\documentclass{emulateapj}

\usepackage[backref,breaklinks,colorlinks,citecolor=blue]{hyperref}        
\usepackage[all]{hypcap}

\newcommand{\rrl}{RR~Lyr\ae}

\shorttitle{Mid-IR Observations of \rrl{} stars in M4}
\shortauthors{Neeley et al.}

\begin{document}


\title{On the distance of the Globular Cluster M4 (NGC 6121) Using RR
  Lyrae Stars: II. Mid-Infrared Period-Luminosity Relations.} 


\author{J. R. Neeley \altaffilmark{1}, M. Marengo \altaffilmark{1},
  G. Bono \altaffilmark{2,3}, V. F. Braga \altaffilmark{2,3}, M. Dall'Ora \altaffilmark{4},
  P. B. Stetson \altaffilmark{5}, R. Buonanno \altaffilmark{6,2}, I. Ferraro \altaffilmark{3},
  W. L. Freedman \altaffilmark{8}, G. Iannicola \altaffilmark{3}, B. F. Madore \altaffilmark{9}, 
  N. Matsunaga \altaffilmark{7}, A. Monson \altaffilmark{9}, S. E. Persson \altaffilmark{9},
  V. Scowcroft \altaffilmark{9}, M. Seibert \altaffilmark{9}} 
  
\slugcomment{Accepted to ApJ: 27 May 2015}





\altaffiltext{1}{Department of Physics and Astronomy, Iowa State University, Ames, IA 50011, USA}
\altaffiltext{2}{Department of Physics, Universit\`a di Roma Tor Vergara, via della Ricerca Scientifica 1, 00133 Roma, Italy}
\altaffiltext{3}{INAF-Osservatorio Astronomico di Roma, via Frascati 33, I-00040 Monte Porzio Catone, Italy}
\altaffiltext{4}{INAF-Osservatorio Astronomico di Capodimonte, Salita Moiarello 16, I-80131 Napoli, Italy}
\altaffiltext{5}{NRC-Herzberg, Dominion Astrophysical Observatory, 5071 West Saanich Road, Victoria BC V9E 2E7, Canada}
\altaffiltext{6}{INAF-Osservatorio Astronomico di Teramo, Via Mentore Maggini s.n.c., I-64100 Teramo, Italy}
\altaffiltext{7}{Kiso Observatory, Institute of Astronomy, School of Science, The University of Tokyo 10763-30, Mitake, Kiso-machi, Kiso-gun, 3 Nagano 97-0101, Japan}
\altaffiltext{8}{Department of Astronomy and Astrophysics, University of Chicago, Chicago, IL 60637, USA}
\altaffiltext{9}{Carnegie Observatories, 813 Santa Barbara Street, Pasadena, CA 91101, USA}

\begin{abstract}

New mid-infrared period-luminosity (PL) relations are presented for \rrl{} variables in the 
globular cluster M4 (NGC 6121). Accurate photometry was obtained for 37 \rrl{} variables using 
observations from the Infrared Array Camera onboard the Spitzer Space Telescope. The dispersion of M4's PL relations is 0.056, and the uncertainty in the slope is 0.11 mag. Additionally, we
established calibrated PL relations at 3.6 and 4.5~\micron{} using published Hubble Space Telescope
geometric parallaxes of five Galactic \rrl{} stars. The resulting band-averaged distance modulus for M4 
is $ \mu = 11.399 \pm 0.007 \textrm{(stat)} \pm 0.080 \textrm{(syst)} \pm 0.015 \textrm{(cal)} \pm 0.020
\textrm{(ext)}$. The systematic uncertainty will be greatly reduced when parallaxes of more stars 
become available from the GAIA mission. Optical and infrared period-color (PC) relations are also 
presented, and the lack of a MIR PC relation suggests that \rrl{} stars are not affected by CO absorption
in the 4.5~\micron{} band.  

\end{abstract}





\section{Introduction}
\label{sec:intro}

\rrl{} variables are important tools in the investigation of many 
fundamental astrophysical problems. They provide crucial constraints on
the physical mechanisms driving radial oscillations and their interplay with 
stellar evolution \citep{cox63, christy66, castor71}. Furthermore, \rrl{} stars
offer the opportunity to study the morphology of the horizontal branch and the 
Oosterhoff dichotomy \citep{oost39}. Current empirical evidence indicates 
that the mean period of fundamental mode (RRab or FU) \rrl{} stars in 
Galactic globular clusters (GGCs hereafter) shows a dichotomous distribution
at 0.55 (OoI) and 0.65 (OoII) days \citep{sandage93}, where the latter group
is more metal-poor. There is also evidence that the dichotomy is the aftermath 
of the hysteresis mechanism suggested by \citet{vanalbada73}, i.e. that the 
pulsation mode depends on the direction of the evolution inside the instability strip
\citep{bono97a, fiorentino12, kunder13}. \citet{baade44} employed the \rrl{} stars 
as a probe to identify the two main stellar populations in the Galaxy, as well as to 
study the stellar content of the Galactic Bulge through low-reddening regions
\citep{baade57,plaut68}.

Beyond stellar evolution, \rrl{} variables have also played a key role in providing 
estimates of cosmological parameters, and have been instrumental in measuring
the distances to a sizable sample of GGcs. This allowed the estimation
of absolute cluster ages, and in turn set the lower limit on the age of the Universe
\citep{hesser91, sandage93, buonanno1998, marinfranch2009, bono2010, 
vandenberg2013}. Moreover, \rrl{} variables have been used to estimate the 
primordial helium content using the A-parameter, i.e. the mass to luminosity 
relation of low-mass central helium-burning stars \citep{caputo83, sandquist2010}. 

\rrl{} stars are also the most commonly adopted Population II distance
indicator. With a lower mass than classical Cepheids, they have the key
advantage to be ubiquitous, and have been identified in both early and
late type stellar systems \citep{vandenbergh1999}. Their
individual distances can be evaluated using multiple diagnostics,
including a visual magnitude-metallicity relation \citep{chaboyer1996,
  bono2003, cacciari2003} and a statistical parallax \citep{kollmeier13,
  dambis13}. More importantly, they obey to well defined near-infrared
(NIR) period-luminosity (PL) relations \citep{longmore86, bono2001,
  bono2003, cat04, bra15}. These PL relations extend to mid-infrared (MIR) bands
where they have the potential of being very accurate distance
indicators due to lower extinction and smaller intrinsic
scatter \citep{mad13, kle14}. The use of the I,V-I reddening free 
period-Wesenheit (PW) relation to estimate the individual distances of \rrl{} stars 
dates back to \citet{sosz2003} and to \citet{majaess2010}. A more recent
theoretical framework developed by \citet{mar15} further supports the use of
optical, optical-NIR, and NIR period-Wesenheit-metallicity (PWZ) relations to
determine individual distances of \rrl{} stars. Empirical validations to the above 
pulsation and evolutionary predictions have been provided by \citet{bra15} for 
\rrl{} stars in the GGC M4 and by Coppola et al. (2015, in 
preparation) for \rrl{} stars in the Carina dwarf spheroidal.

The Carnegie \rrl{} Program (CRRP) aims to take full advantage of the
unique characteristics of these stars in order to reduce the remaining
sources of uncertainty in the Hubble constant to $\pm 2$\%. \rrl{} MIR PL relations
will be used as the foundation of an independent Population II cosmological 
distance scale to calibrate TRGB distances for nearby distances, which in turn can
be used to calibrate Type Ia Supernova distances. To achieve this goal, we 
observed over 1,700 \rrl{} variables in 31 selected GGCs, as well as $\sim 3,000$ \rrl{} stars in
strategically distributed areas in the Galactic Halo and the Bulge. In
addition, we observed 48 of the nearest, brightest and less reddened \rrl{} stars
intended to be used as zero point calibrators. These observations have
been conducted during the warm mission of the Spitzer Space Telescope
\citep{werner04} Infrared Array Camera \citep[IRAC,][]{fazio04}, at
3.6 and 4.5~\micron{} wavelength, with a cadence designed to obtain
complete coverage of each \rrl{} variable over at least one full
period.

In this work we focus on the GGC Messier 4 (M4, NGC~6121). This
cluster is an ideal laboratory for stellar population studies given
its proximity to the Sun, which allows us to obtain accurate photometric
and spectroscopic data for member stars well below the main-sequence
turnoff. Due to these characteristics, M4 has been the subject of
intensive observational campaigns over a wide range of wavelengths. It
has a well characterized differential extinction of $E(B - V) = 0.37 \pm 0.10 $ mag,
where the uncertainty is the dispersion due to differential reddening \citep{hen12}, and mean metallicity
of $[\textrm{Fe}/\textrm{H}] = -1.10$, \citep[and references
therein]{bra15}. In support of the CRRP program, we have analyzed
available multi-epoch optical and NIR data, leading to the
identification and characterization of 45 \rrl{} variables
\citep{ste14}. From these observations, we have derived accurate
distance moduli based on optical and NIR PL and PW relations
\citep{bra15}. In this paper we combine our previous results with the
new MIR data obtained as part of the CRRP campaign.

In Section~\ref{sec:obs} we present our new Spitzer photometry.
Light curves for all the \rrl{} variables in our sample are measured in
Section~\ref{sec:mags}. In Section~\ref{sec:PL} we derive MIR PL and
period-color (PC) relationships for the cluster \rrl{} variables,
while in Section~\ref{sec:DM} we calculate the M4 distance modulus by
calibrating our PL zero point using five nearby calibrator \rrl{} stars with
known parallax, also observed as part of the CRRP program. Dependence
of the PL zero point from metallicity is also discussed in
Section~\ref{sec:DM}, while Section~\ref{sec:concl} summarizes the
results of this work.

\section{Observations, Data Reduction and Photometry}
\label{sec:obs}

The results of our ground-based optical and NIR monitoring of the
stellar population in M4 have been published in \citet{ste14} and
\citet{bra15}. In this paper we extend our wavelength coverage to the
MIR, by including multi-epoch 3.6 and 4.5~\micron{} photometry of a
large portion of the cluster. 

\begin{figure*} \includegraphics[angle=0,scale=.75]{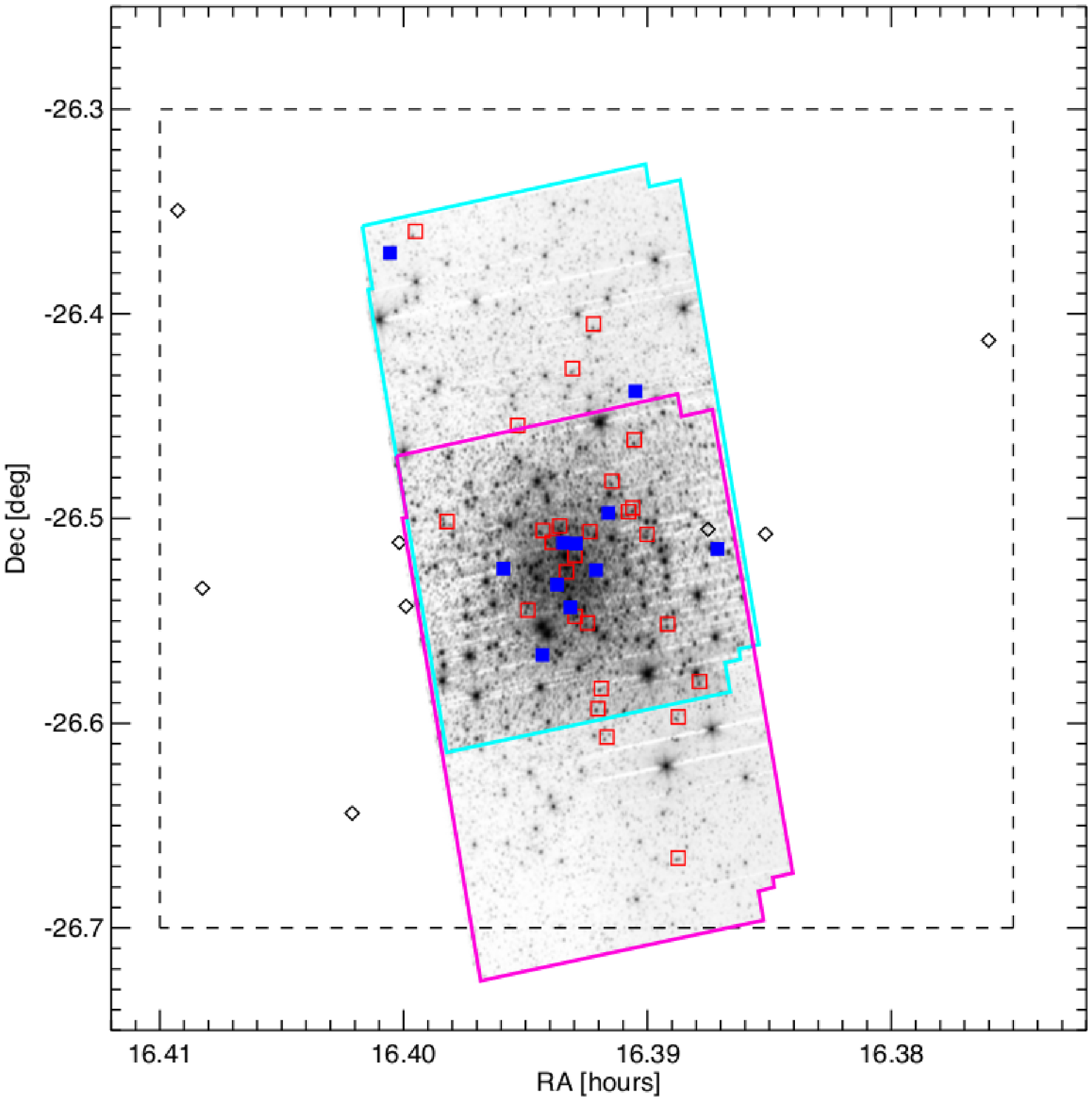}
  \caption{Map of all M4 stars for which we have MIR photometric
    measurements. The open red and filled blue squares display the position of  
    fundamental and first overtone \rrl{} stars respectively. 
    The black diamonds are \rrl{} from our optical catalog that are not
    matched in the IRAC catalog. The solid cyan and
    magenta lines indicate the coverage of the IRAC 3.6 and 4.5
   \micron{} data, respectively. The dashed line indicates the
    approximate field of view of the optical photometry in
    \protect\citet{ste14}. North is up, east is left.}\label{fig:map} \end{figure*}

Our IRAC observations of M4 were executed on 2013 June 2 - 3 as part
of the warm Spitzer Space Telescope Cycle~9 (PID 90088). The cluster
was observed at IRAC 3.6 and 4.5 $\micron$ bands at 12 equally spaced epochs
over $\sim 14$ hours. For each epoch we obtained 30 frames with a
medium scale five-point Gaussian dither pattern and a $3 \times 2$
mapping, taken using the IRAC full frame mode with a frame time of 30
seconds to provide a median total exposure of 804~seconds per pixel.
An $11' \times 9'$ area at the center of the cluster was covered in
both the 3.6 and 4.5~\micron{} bands. Due to the IRAC focal plane
array configuration, in the outer regions of the cluster only one band
is available, with the area just north of the center of the cluster
only covered at 3.6~\micron{} and the region south of the center only
covered at 4.5~\micron. The total area mapped by our observations
covers a strip of $11' \times 22'$, roughly cutting North-to-South
through the center of the cluster, as shown in Figure~\ref{fig:map}.
This is about one third of the total cluster area that was mapped in
our ground-based optical and NIR observations, but is still $\sim 1.5
\times$ the size of the half-mass radius of M4 ($3 \farcm 65$
according to \citealt{harris1996}).


The CRRP IRAC data were reduced starting from the Basic Calibrated
Data (BCDs) generated by the Spitzer IRAC pipeline version S19.1.0.
Photometry was performed using the most recent version of the
DAOPHOT/ALL\-STAR/ALLFRAME suite of programs for crowded-field stellar
photometry \citep{stetson1987, stetson94}. In order to mitigate the effect of
blending in the photometry of the M4 crowded field, we used an input source 
catalog derived by the higher angular resolution optical and NIR images described
in \citet{ste14}. In addition to higher resolution, this input source catalog is also 
deeper. For the optical and NIR, accurate photometry was achieved down to 24
and 22 magnitudes respectively, compared to 17 mag for the MIR images. 
This catalog was matched to the Spitzer data after applying higher 
order geometric corrections to the individual IRAC BCDs than the default ones 
provided by the Spitzer IRAC pipeline. With this procedure we matched 17,595 
sources at 3.6~\micron{} and 16,313 sources at 4.5~\micron, for
which we measured an instrumental magnitude with variable Point Spread
Function (PSF) fitting. The astrometric accuracy of our final source
catalog is of the order of $0.3''$. 

For the optical and NIR images, good photometry was achieved down to 24 and 20 magnitudes respectively

The photometry was calibrated to standard IRAC Vega magnitudes using
aperture photometry on mosaic images. The mosaics were generated using
the IRACproc package \citep{sch06}, a Perl Data Language wrapper
script for the standard Spitzer Science Center mosaicking software
MOPEX \citep{mak05}. We adopted a mosaic pixel scale of 0.61$''$/pix,
or half the native IRAC pixel scale. The stars we selected for
aperture photometry calibration are far from the center of the cluster
to reduce blending. In addition, we removed any possible variable
stars, by avoiding sources with a Welch \& Stetson variability index 
\citep{wel93, ste96} greater than $1\sigma$ from the median. We selected a
total of 38 stars at 3.6~\micron{} and 44 at 4.5~\micron. We used an
aperture of 3 native IRAC pixels ($\simeq 3.66''$) with an annulus of
3 to 7 pixels ($3.66''$ to $8.54''$). The aperture corrections
necessary to convert magnitudes derived with this aperture/annulus
combination to the standard IRAC 10 pixel aperture photometry are
1.1284 and 1.1274 for 3.6 and 4.5 $\micron$ respectively, calculated
on isolated stars in the mosaics. The calibrating stars cover a range
of 4 magnitudes and are distributed randomly across the array, outside
of the central region.

Despite the relatively small size of our adopted photometric
apertures, the crowding of the field even outside the center of the
cluster (M4 is projected behind the Scorpius-Ophiuchus cloud complex)
still results in blending contamination of aperture photometry for our calibration sources.
These blended sources are however in many cases resolved by the PSF
fitting procedure performed in DAOPHOT on the higher resolution
visible and NIR images, from which our input catalog is derived. We
thus used our point source photometric catalog to improve the photometric
zero point calibration by subtracting from the calibrator aperture photometry
the flux (measured from the PSF-fitting photometry) of any blended
sources found within our chosen aperture. For each blended source we
estimated the contaminating flux by measuring the fraction of its flux
falling within our 3 pixel aperture, as a function of its centroid
distance from the aperture's center. For this task we used the IRAC
PSF described in \citet{hor12}, taking advantage of its higher
sampling ($\sim 1/100$ of the native IRAC pixels) to derive
corrections for fractional pixel shifts. At 3.6~\micron{} 17 of our 38 calibrators 
were found to be contaminated by an average of 0.03 mag. The 4.5~\micron{} 
region was less crowded, resulting in only 7 out of 44 contaminated calibrators, 
but contaminated by 0.11 mag on average. 

The magnitudes of the de-blended calibrators were then subtracted from
their DAOPHOT instrumental magnitudes to derive the final calibration
zero point. We found no trend in the calibration zero point with either brightness
or array location. The uncertainty in our final calibration is 0.018
mag at 3.6~\micron{} and 0.017 mag at 4.5~\micron{}, based on the
standard deviation between the instrumental and measured magnitudes.
This standard deviation of the calibration increases dramatically for
the fainter stars in our calibration sample. If we select only
calibrating stars brighter than 14 mag, then the error in the
calibration is reduced to 0.015 at 3.6~\micron{} and 0.013 at
4.5~\micron.

At the end of this process, we obtained a photometric catalog in both
IRAC bands for all point sources detected in our field of view. From
this catalog we matched 37 of the 45 \rrl{} stars with optical photometry
derived by \citet{ste14}, 26 of them fundamental mode (FU, RRab) pulsators, and 11
pulsating in first overtone (FO, RRc). Of the remaining eight \rrl{} stars in the
cluster, seven are outside the area mapped in at least one of the two IRAC
bands, and one (V64) is a field variable in the background of the
cluster, too faint to be detected by IRAC. All M4 \rrl{} stars are indicated
in Figure~\ref{fig:map} by squares, with open red and filled blue symbols for
FU and FO \rrl{} stars detected in at least one IRAC band.

The final magnitude of each non-variable star was derived by averaging
the value measured in all epochs and in all dithers. For the \rrl{}
variables we instead preserved the photometry for each epoch,
averaging only on the individual dithers, after correcting the
photometry in each dither for the warm Spitzer location-dependent
photometric corrections available at the Spitzer Science Center web
site.\footnote{\url{http://irsa.ipac.caltech.edu/data/SPITZER/docs/irac/calibrationfiles/locationcolor/}}
The photometric uncertainty in each epoch is given by the standard
deviation of all of the available measurements. A sample of the individual data points for one star, V1,
is given in Table~\ref{tab:phot}; all data is available in the online version of this article.
 A few \rrl{} stars were covered in only one of the five dither positions, and one magnitude measurement was available in each epoch. For these stars (V1, V42, V2, and V29) the photometric 
uncertainty is instead derived from the \emph{repeatability} parameter provided 
by DAOPHOT, which tends to be larger than the photometric error based 
exclusively on photon and background noise.

\section{\rrl{} Light Curves and Average Magnitudes}
\label{sec:mags}

Our MIR sample of M4 \rrl{} variables consists of 37 stars, whose
location is shown in Figure~\ref{fig:map}. Of those, 26 are pulsating
in fundamental mode (FU, RRab) and 11 are first overtone pulsators (FO, RRc). Not all
sources have complete coverage in the IRAC dataset: of the
37 \rrl{} stars in the area mapped by at least one IRAC band, only 28 have
photometry at both 3.6 and 4.5~\micron, with six sources covered only
at 3.6~\micron{} and three covered only at 4.5~\micron.
The periods of our \rrl{} stars range between 0.2275~days
(V49) to 0.6240 days (V39), plus a long-period FU \rrl{} (V52) with a
period of 0.8555 days.

\begin{figure*}
\includegraphics[angle=0,scale=.89]{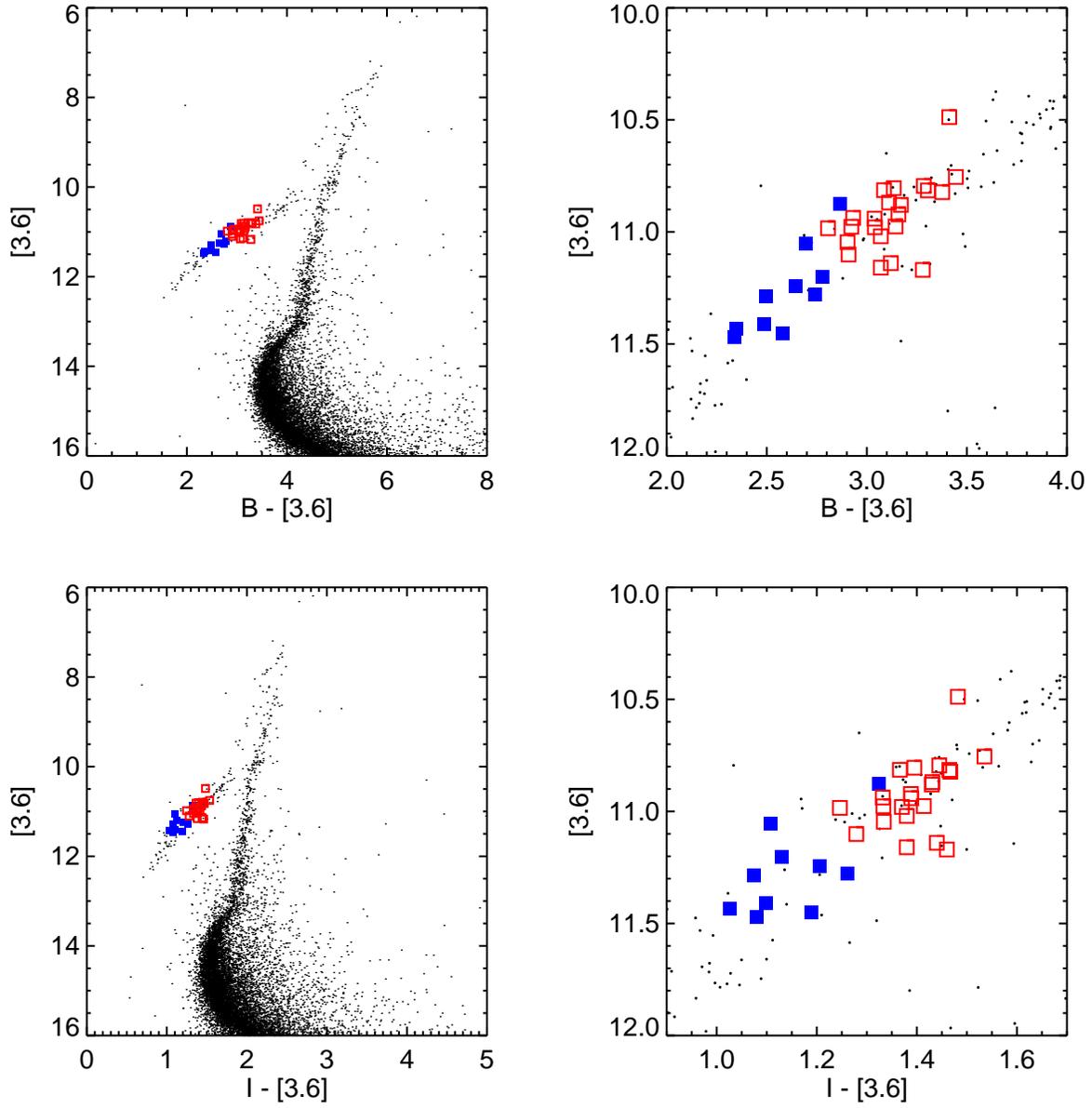}
\caption{Sample color-magnitude diagrams for all sources detected in
  the IRAC 3.6~\micron{} band, with $B$ and $I$ photometry from
  \protect\citet{ste14}. Fundamental and first overtone \rrl{} variables are
  represented by open red and filled blue squares, respectively. The right panels 
  show a close up of the horizontal branch.}\label{fig:CMD} 
\end{figure*}

Figure~\ref{fig:CMD} shows representative color-magnitude diagrams
(CMDs) for M4 based on our MIR and optical photometry, with the IRAC
3.6~\micron{} magnitude vs. the $B - [3.6]$ (\emph{top left}) and $I - [3.6]$
(\emph{bottom left}) colors. 
Note how the high photometric quality of our data allows a well characterized turn-off in all color combinations, 
and a well populated sub-giant, RGB and AGB branch. The FO and FU \rrl{} stars
are represented by filled blue and open red squares respectively. We see a clear
separation of the FO and FU \rrl{} stars in the horizontal branch of the
CMD, with the FO bluer than the FU variables (\emph{right panels}).

\begin{figure*}
\includegraphics[angle=0,scale=.9]{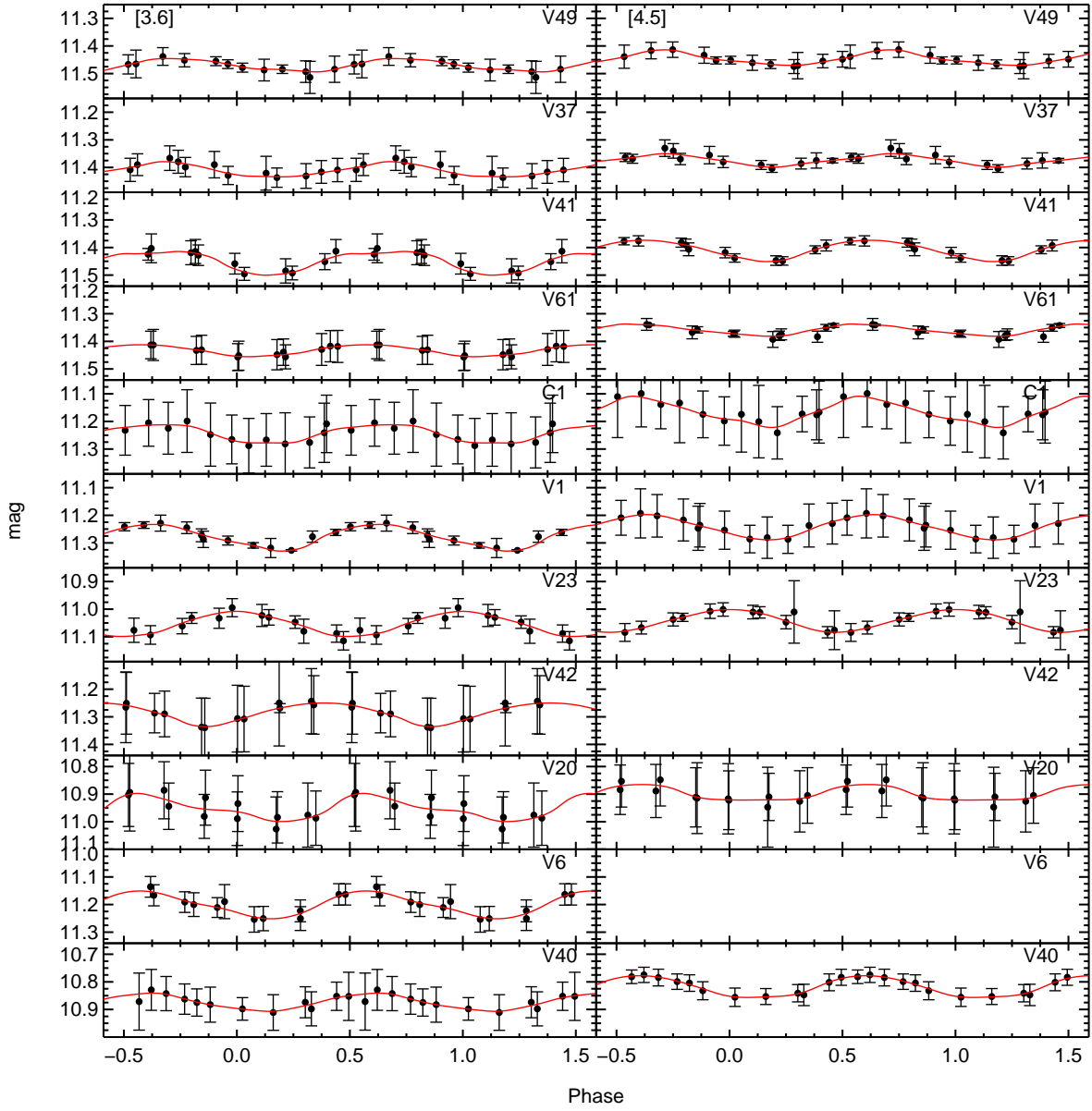}
\caption{IRAC light curves for all FO \rrl{} variables in M4.
  The light curves are arranged from top to bottom in order of
  increasing period. The left and right panel display the light curves
  at 3.6 and 4.5~\micron. The solid red line is the smoothed light
  curve generated using a Gaussian local estimator algorithm (see text
  for more detail). All light curves cover the same range in magnitude. }\label{fig:FO-lightcurves}
\end{figure*}

\begin{figure*}
\includegraphics[angle=0,scale=.9]{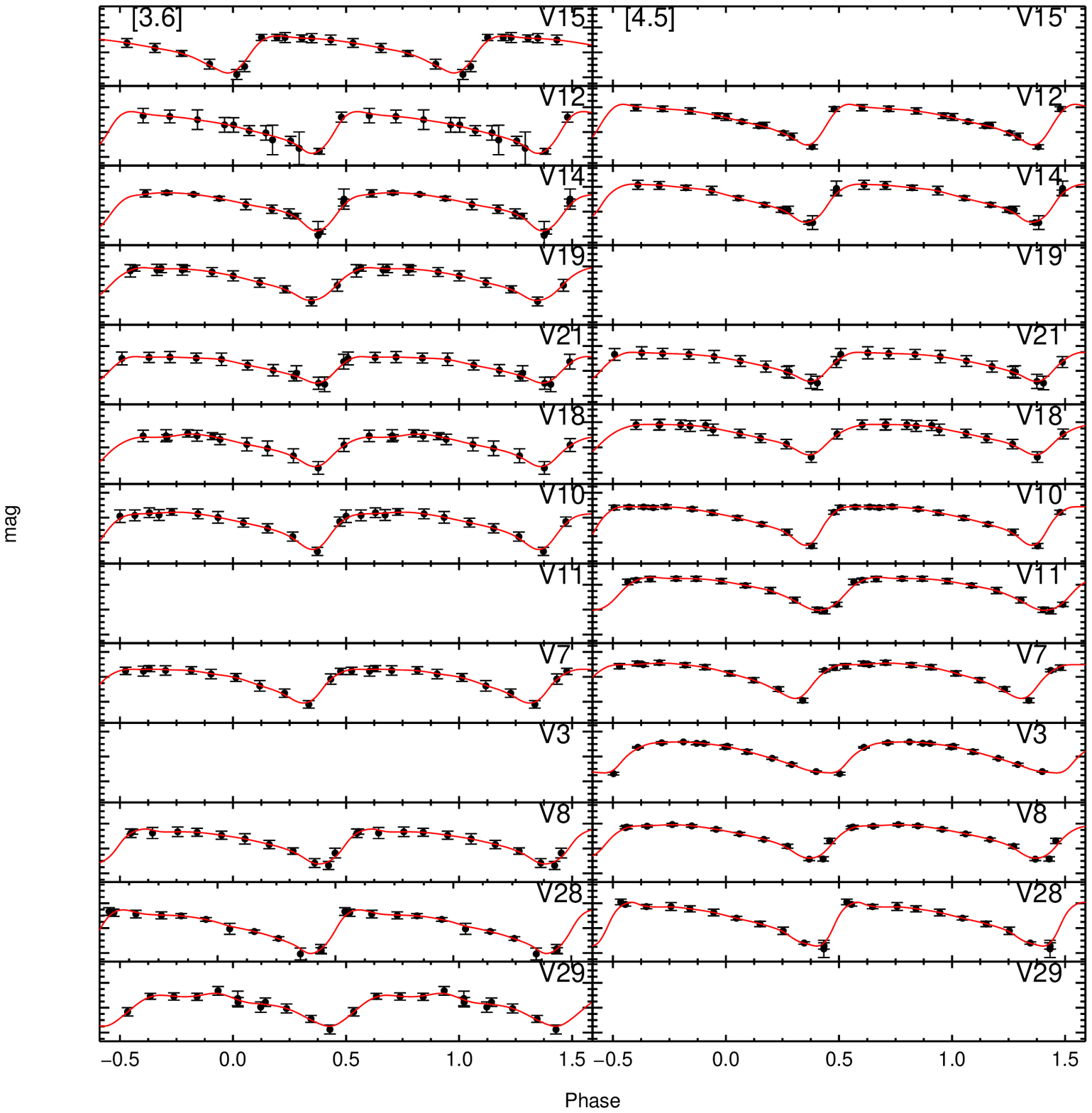}
\caption{Same as Figure~\ref{fig:FO-lightcurves} but for all FU
  pulsators observed by IRAC.}\label{fig:FU-lightcurves}
\end{figure*}

We derived MIR light curves for each \rrl{} in our catalog by phasing
their multiple-epoch photometry based on the epoch of maximum ($T_0$)
and the period derived by \citet{ste14} using the Lomb-Scargle method
from the visible data. Sample light curves are shown in
Figure~\ref{fig:FO-lightcurves} for FO variables and \ref{fig:FU-lightcurves} for 
FU variables. The light curves are shown in order of increasing period. 
Following \citet{mad13}, for each star we
derived a smoothed light curve using a Gaussian local estimation
(GLOESS) algorithm, a method where a second-order polynomial is fit
locally and points are weighted by their Gaussian distance from an
interpolation point. A full description of the GLOESS method is given
in \citet{per04}. The mean magnitude for all stars was
then determined by integrating the smoothed light curve flux intensity
over one period, then converting the result back into a magnitude. The
variability amplitude in magnitude was also derived from the smoothed
light curve, and the uncertainty is calculated using only the data points that 
lie within the Gaussian window of the maximum and minimum of the smoothed
curve.

The uncertainty in the mean magnitude is calculated by the sum in quadrature of 
photometric uncertainty of each observation ($\sqrt{\Sigma \sigma_i^2} / N$) and 
the uncertainty in the fit ($A / (M \sqrt{12})$) where $N$ is the number of 
observations, $\sigma_i$ is the photometric uncertainty, $A$ is the
peak-to-peak amplitude, and $M$ is the number of uniformly spaced points
\citep[following][]{sco11}. The value of $M$ ranges from five to twelve; 
five (e.g. V41) when our observations cover more than two periods and twelve 
when our observations cover one complete period (e.g. V5). Some light curves 
however were randomly sampled (e.g. V52, where the observations do not cover 
a full period), and the uncertainty in the fit is instead $A / \sqrt{12N}$ where N is 
the total number of observations. Table~\ref{tab:Smean} gives the 3.6 and
4.5~\micron{} IRAC mean magnitudes and amplitudes for the 37 \rrl{} stars in
the IRAC field. 

Figure~\ref{fig:bailey} shows the luminosity amplitude vs. period (Bailey
diagram) for all \rrl{} stars: filled blue squares for FO and open red squares 
for FU pulsators. Black crosses indicate FU \rrl{} stars
identified in \citet{ste14} as Blazhko variables (the Blazhko effect
is characterized by a conspicuous modulation of amplitude and phase;
\citealt{blazko1907}). Recent long-term monitoring of Blazhko \rrl{} by 
the Kepler satellite has provided evidence of not only period-doubling
\citep{szabo2010} but also the possible occurrence of additional periodicities 
on a time-scale longer than the typical Blazhko period \citep{kolen2011, chadid2011}.
As expected the FO and FU \rrl{} variables are well
separated in both period and amplitude, and the amplitude of the FU
pulsators show a clear trend of decreasing with period.
Figure~\ref{fig:AP} shows instead the amplitude ratio between optical
bands and the IRAC 3.6~\micron{} band. The amplitude 
ratio plot can be used as a diagnostic for blends of stars with different color
temperature. Two FO \rrl{} stars (V40
and C1) show an unusually high optical to MIR ratio, possibly
indicative of the presence of a faint optically-bright (hot) companion
or blend. Another FO \rrl{} (V49) shows instead an unusually low
amplitude ratio, possibly the consequence of having an infrared-bright
(cool) companion (or blend). It is also interesting to note that
several (four out of seven at 3.6~\micron{} and four out of six at 4.5~\micron) of
the FU \rrl{} stars that lie below the general amplitude ratio trend are
Blazhko variables, possibly due to these stars being observed at
different phases in their amplitude modulation cycle (the optical and
MIR data were not acquired simultaneously). If this is the case, the
remaining FU \rrl{} stars in the same amplitude ratio group (V5, V3, and 
V52) could also be Blazhko stars that were missed by the period studies 
(or they could be stars with a faint cool companion or blend).

\section{Mid-IR Period-Luminosity and Period-Color Relationships}
\label{sec:PL}

The left panels of Figure~\ref{fig:PL-warm} show the IRAC PL
relation for FO (filled blue squares) and FU (open red squares) \rrl{} stars in M4.
Blazhko variables are indicated by black crosses. The statistical
uncertainties of the mean magnitudes are smaller than the size of the
plotted symbols, and are not shown. Residuals from the PL relation were used as a diagnostic for blended variables. The central regions of M4 are not particularly dense, and no clear cutoff in 
radial distance from the center was apparent. Instead we employed a threshold in sigma 
to identify candidate blends. Two stars, V20 and V21, in the central regions of the cluster 
fall outside the threshold. The variable V20 was over $3\sigma$ above the fitted PL relation, 
and as shown in Figure~\ref{fig:FU-lightcurves}, has a very poor quality light curve. The 
average magnitude of V21 was  more than $5\sigma$ above the fitted relation. These stars 
were not included in the final fit of the PL relations, and are shown in Figure~\ref{fig:PL-warm}
as open circles.

All average magnitudes are corrected for extinction; albeit MIR extinction corrections are small
compared to optical wavelengths, they still can not be ignored given
the high photometric accuracy provided by IRAC. We used the
total-to-selective extinctions of $A_{[3.6]}/E(B-V) = 0.203$ and
$A_{[4.5]}/E(B-V) = 0.156$ from \citet{mon12} and a color excess of
$E(B-V) = 0.37\pm0.10$ from \citet{hen12}. This results in extinction
corrections of $0.075\pm0.020$ and $0.058\pm0.016$ mag at 3.6 and 4.5~\micron{},
respectively.

We calculated the zero point and slope (with their uncertainties and
the best fit standard deviation) for the PL relation using an unweighted 
least squares fit. An unweighted fit is preferred to avoid biasing by brighter 
longer period variables, which tend to have smaller photometric uncertainties. 
The PL relations take the form 
\begin{eqnarray}
m=a+b(\log(P)+0.55)\\
m=a+b(\log(P)+0.26)
\end{eqnarray}
where $\log(P) = - 0.55$ and $\log(P) = - 0.26$ are representative of the mean 
period of the FO and FU variables respectively. 
The scatter between the FO and FU PL relations is
comparable. Due to the small number of FO variables, we also derived
the global PL relations for all \rrl{} stars (Figure~\ref{fig:PL-warm}, \emph{right}), having
\emph{fundamentalized} the periods of the FO variables using the
relation $\log (P_{FU}) = \log(P_{FO}) + 0.127$ \citep{iben71, rood73, cox83}. 
The fundamentalized PL relation takes the form 
\begin{eqnarray}
m=a+b(\log(P)+0.30)
\end{eqnarray}
where $\log(P)=-0.30$ is representative of the mean fundamentalized period of all \rrl{} stars in our sample.
The zero point $a$, slope $b$, their errors, and the standard deviations for all PL 
relations are given in Table~\ref{tab:PL}. Note how the secondary modulation 
of the candidate Blazhko variables seem to have no measurable effect on 
their average magnitudes, as they fit well on the PL relation.

Figure~4 in \citet{mad13} showed that, using available data, the slope
of the \rrl{} PL relation is a monotonically increasing function of
wavelength, and asymptotically approaches a wavelength-independent
slope. In Figure~\ref{fig:PL-parameters}, we recreate this plot by
adding the optical and near-infrared slopes from \citet{bra15}, the
IRAC MIR slopes calculated above, and the results from \citet{kle14}.
Our data is consistent with the values found by \citet{mad13} and
\citet{kle14} for the WISE W1 and W2 bands, and confirm with a high
level of accuracy that the slope approaches the value of $-2.60$ at
infrared wavelengths, as predicted by the period-radius relation in
\citet{burki86}.

\subsection{Period-Color Relationship}
\label{ssec:PC}

Figure~\ref{fig:PC} shows optical and infrared period-color (PC)
relations for all M4 \rrl{} variables observed with IRAC (as before,
filled blue and open red squares indicate fundamentalized FO and FU). 
We fit the data with an unweighted least squares procedure;
the best fit parameters are listed in Table~\ref{tab:PC}. 
The figure clearly shows how the slope of the PC relation becomes more
shallow as the distance between the two passbands decreases, as
expected if the color is a temperature effect related to the radius of
the pulsator. The $V - K$ PC relation, in particular, has been singled
out for being the least affected by temperature uncertainties when
deriving \rrl{} distances with the Baade-Wesselink method
\citep{carney1992, cacciari2000, mcnamara2014}, as opposed to using
the standard $B - V$ color. Our figure shows that PC relationships
using MIR colors (and the IRAC 3.6~\micron{} magnitude in particular)
are even more advantageous, thanks to the steeper slope (i.e. $\simeq
1.93$ for the $V - [3.6]$ relation vs. $\simeq 1.87$ in the $V - K$
fit) and slightly smaller dispersion around the best fit relation.

The last panel in Figure~\ref{fig:PC} shows the IRAC $[3.6] - [4.5]$ 
PC relation. In this case the slope is very shallow ($0.046 \pm
0.033$), consistent with zero within $\sim 1.4 \sigma$. The scatter
around the best fit PC relation is also quite large (0.017~mag),
three times larger than the uncertainty in the average magnitudes of
individual stars. The lack of a significant $[3.6] - [4.5]$
PC relation for \rrl{} stars suggests that these stars do not develop
variable CO molecular absorption in their atmosphere, which is
observed in other pulsating stars in the same instability strip (e.g.
classical Cepheids). \citet{sco11} found a well defined $[3.6] - [4.5]$ PC
relation for Large Magellanic Cloud Cepheids with $\log P \la 2$. The
negative slope of the PC relation and blue colors of these stars is
explained by the dissociation and recombination of CO molecules in the
IRAC 4.5~\micron{} band, as previously noted by \citet{marengo2010}.
This effect is particularly strong in Cepheids with $P > 10$~days,
and is reflected in a precise phasing of the light and color curve of
these stars. Shorter period Cepheids are instead characterized by a
flat color curve and red colors. This is further supported by \citet{mon12},
which found that CO absorption drops off for temperatures greater than 6000 K. 
Given that these \rrl{} stars have higher effective temperatures, earlier spectral types and lower
metallicity than Cepheids, it is not surprising that we do not find
strong CO effects in the PC relation and color curve. The $[3.6] - [4.5]$ 
PC relation appears to be flat, even when including Galactic \rrl{} stars with very 
different metallicities. Therefore, \rrl{} observations at 4.5~\micron{} are not
limited by strong metallicity effects as Cepheids are, and this
band can be successfully used for distance scale measurements with
similar accuracy as the 3.6~\micron{} band. The weakness of the
4.5~\micron{} CO band in \rrl{} stars, however, prevents this feature from
being used as metallicity indicator for this class of variables.

\section{MIR PL Zero Point Calibration and M4 Distance Modulus}
\label{sec:DM}

The CRRP project relies on establishing \rrl{} variables as highly
accurate indicators for the first rung of the cosmological distance
scale, by using their MIR PL relation. The zero point of this
relation, however, needs to be calibrated. Efforts in this direction
have been made in the recent past by using Galactic \rrl{} stars observed by
WISE \citep{mad13, kle14}, but the accuracy of the calibration has
been limited by the uncertainties in the photometry and distance of
the calibrators. As mentioned in Section~\ref{sec:intro}, the CRRP
project addresses this issue with the observations of 43 Galactic
\rrl{} stars whose geometric parallax will be determined better than 2-3\%
with Gaia, and five bright \rrl{} stars with parallax already determined
by \citet{ben11} using the Hubble Space Telescope (HST) Fine Guidance
Sensor (FGS). Ground-based monitoring and spectral observations of the
calibrators is ongoing, to ensure the characterization of their
pulsation properties and, if required, to allow the calibration of the
metallicity dependence of the PL relation.

The observation of a large sample of GGCs (themselves used as distance
indicators) as part of the CRRP program provides a separate avenue to
test the calibration of \rrl{} stars and probe for metallicity effects. The
distance of M4, in particular, has been the subject of intense analysis,
most recently by \citet{bra15} by using a theoretical and empirical
calibration of the \rrl{} optical and NIR PL relation for the cluster.
The distance modulus derived in this work can be used as an alternative
calibration of the MIR PL zero point, as well as to test the effects
of metallicity on the zero point derived from the five HST/FGS \rrl{}
calibrating stars. A similar approach will be followed for the remaining 30 GCCs 
in the CRRP program (spanning a wide range in metallicity), for which 
independent distance moduli will be derived in the optical, NIR and MIR 
from their population of \rrl{} stars.

\subsection{Zero Point Calibration Using HST \rrl{} Stars}
\label{ssec:GC}

As mentioned above, at present there are only five \rrl{} variables with
available geometric parallax, i.e. the aforementioned stars observed
with HST/FGS by \citet{ben11}: RR~Lyr, UV~Oct, SU~Dra XZ~Cyg (FU
pulsators) and RZ~Cep (FO). The four FU pulsators are the same stars
used by \citet{mad13} to calibrate the \rrl{} PL relation in the WISE
bands.  Even for these stars, however, the uncertainty in their
distance modulus is still large (up to 0.25~mag), significantly
reducing their effectiveness as calibrators for the distance
scale. All five stars have been observed as part of the CRRP program
in order to reproduce the \citet{mad13} zero point calibration using
the IRAC 3.6 and 4.5~\micron{} bandpasses, and with better phase 
sampling and much smaller ($\sim 10\times$ better S/N ratio)
photometric error. Their basic properties (derived from
\citealt{ben11}) are listed in Table~\ref{tab:gal}: their periods range from 
0.3086 to 0.6642~days, and their [Fe/H] ranges from
$-1.43$ to $-1.83$, significantly more metal-poor than M4 itself. A detailed 
analysis of the calibration of the zero point using these five \rrl{} stars at multi--wavelengths 
will be presented in a forthcoming paper. 

The mean magnitude of each star was computed as explained in Section~\ref{sec:mags}. Each
calibrator was then corrected for extinction, adopting the $E(B - V)$
reddening from \citet{ben11}. The extinction corrections were computed
using the same $A_{3.6}/E(B-V)$ and $A_{4.5}/E(B-V)$ relations as in
Section~\ref{sec:mags}. The absolute mean magnitudes, as well as the 
parallax, LKH correction \citep{lk73,han79}, distance moduli, extinction corrections, and apparent mean magnitudes
are given in Table~\ref{tab:gal}. \citet{ben11} contains two conflicting results about the parallax 
to RZ Cep and although the value of 2.12 mas is their preferred solution (Benedict, private
communication), we chose to adopt the 2.54 mas solution with an $A_{v}$ of 1 (assuming a mean
spectral type of A5 \citep{preston59} and $(B-V)_{0} = 0.15$) which appears to be more
consistent with the new photometric data. The dominating factor in the uncertainty of the 
absolute mean magnitudes is the uncertainty in their distances.

Based on theoretical NIR PL-metallicity (PLZ) relations, the slope of
the \rrl{} PL relation is expected to be only weakly dependent on
metallicity \citep{bra15}. Therefore, for this work we elected to adopt the well
constrained slope of the fundamentalized PL relation derived in
Section~\ref{sec:PL}, rather than relying on just five Galactic
calibrators (having a smaller range in period than the M4 \rrl{} stars) to
simultaneously fit both slope and zero point. Figure~\ref{fig:HST-fit}
shows the resulting PL relation, derived after fundamentalizing the
period of RZ~Cep. The calibrated PL relations derived using a one 
parameter fit are given by

\begin{eqnarray}
M_{[3.6]}= -2.332(\pm0.106)\log(P) - 1.176(\pm0.080) \\
\nonumber \sigma= 0.095 \\ 
M_{[4.5]} = -2.336(\pm0.105)\log(P) - 1.199(\pm0.080) \\
\nonumber \sigma= 0.095 
\end{eqnarray}
\noindent
where the uncertainty in the zero point is calculated by $\sqrt{\sum\limits_{i} \sigma_i^2 }/5$
and $\sigma_i$ are the uncertainties of the absolute magnitudes for the five calibrator stars 
used in the fit. The largest factor in the uncertainty of the zero point derived above is due to the
uncertainty in the distance modulus of the calibrators, but it is a factor of 3 better than the
value obtained in \citet{mad13} where both the slope and zero point were fit simultaneously and only four stars were used. 
In both IRAC bands, the slope we obtain is marginally shallower (but still consistent, within the error) than the 
values published by \citet{kle14} and \citet{mad13} for the WISE W1 and W2 bands.
Further improvements in the uncertainty of the zero point will be possible when more accurate
parallaxes for all our calibrator \rrl{} stars are obtained with the Gaia mission.

\subsection{M4 Distance Modulus}
\label{ssec:M4-DM}

From the zero point calibration of the MIR \rrl{} PL relation in the
previous section, we can derive the true distance modulus of M4:
 
\begin{eqnarray} 
\mu_{[3.6]} = 11.406 \pm 0.010 \, (\textrm{stat}) \pm 0.080 \, (\textrm{syst}) \\
\nonumber \pm 0.015  \, (\textrm{cal}) \pm 0.020 \, (\textrm{ext}) \\ 
\mu_{[4.5]} = 11.391 \pm 0.010 \, (\textrm{stat}) \pm 0.080\, (\textrm{syst}) \\
\nonumber \pm 0.013 \, (\textrm{cal}) \pm 0.016 \, (\textrm{ext}) 
\end{eqnarray}

\noindent
where we have adopted the M4 PL relation using the fundamentalized
period of FO pulsators. The error on
the distance modulus comes from four sources. The \emph{statistical}
error is the uncertainty in the zero point of the cluster PL relation,
and is derived from the least squares fit to the IRAC data.
The \emph{systematic} error is the uncertainty in the calibrated PL
zero point, given in the previous section. The \emph{calibration} error is the dispersion in our
photometric zero point calibration to standard IRAC Vega magnitudes,
described in Section~\ref{sec:obs}. The \emph{extinction} error is the derived
from the uncertainty due to differential reddening given in Section~\ref{sec:PL}.

The distance modulus we find is slightly larger than expected, based on
previous works ($11.30 \pm 0.05$ in \citealt{kal13}, $11.28\pm0.06$ in
\citealt{hen12}, and $11.35 \pm 0.03 \pm 0.05$ in \citealt{bra15},
this last value derived using RR~Lyr itself as zero point calibrator),
but agree within $1 \sigma$. If we exclude the systematic error, which
will be significantly reduced once high accuracy geometric parallaxes
for a large sample of calibrators will become available from Gaia, we
have an overall uncertainty better than 0.5\% in each band. The average
distance modulus of the two wavelengths then is:

\begin{eqnarray}
\mu = 11.399 \pm 0.007 \, (\textrm{stat}) \pm 0.080 \, (\textrm{syst}) \\
\nonumber \pm 0.015 \,(\textrm{cal}) \pm 0.020 \, (\textrm{ext})
\end{eqnarray}

\noindent
where the statistical error is reduced by a factor of $\sim \sqrt{2}$
(since by using both bands we take advantage of twice the number of
available datapoints). Note that in this case the overall error is reduced 
to 1\%, small enough to allow the study of higher order parameters in 
the \rrl{} PL relation, first of all the possible dependence on metallicity, 
once data from more clusters is analyzed. 

Based on the models presented in \citet{bra15}, while the dependence
of the PL slope from metallicity is weak, we do expect a larger effect
of [Fe/H] on the PL relation zero point. Therefore calibrating the PL
relation with Galactic \rrl{} stars, all with different metallicities, may
induce a significant error. Figure~\ref{fig:met} addresses this issue,
by plotting the residual between the absolute mean magnitude of each
calibrator and the value predicted by the PL relation, as a function
of their metallicity. The metallicities used in the plot are listed in
Table~\ref{tab:gal}, and are the same listed in \citet{ben11} (see
references therein for the original source of the metallicity
measurement), converted from the original \citep{zw84} metallicity
scale to the more recent \citet{car09} scale. For RR~Lyr and UV~Oct an
extra step was necessary to correct for the different solar abundances
used by \citet{kol10} when measuring the metallicity of these two
stars, with respect to the \citet{car09}. Figure~\ref{fig:met} also plots the
difference between the M4 distance modulus calculated above for M4,
and the value of $11.35 \pm 0.03 \pm 0.05$ derived by \citet{bra15}
from NIR PL relations calibrated using RR~Lyr. Due to the large uncertainties 
in absolute magnitude and metallicity of the Galactic \rrl{} stars, no trend is apparent 
(slope of -0.2$\pm$0.3 mag/dex).

If we adopt the M4 distance modulus value of $11.283 \pm 0.001 \pm
0.018$ derived by \citet{bra15} from NIR data and
\emph{theoretical} PLZ relationships for \rrl{} stars (to avoid the large
uncertainty in the calibrator distance that is affecting the zero
point calibration), we obtain the following calibrated \rrl{} PL
relations for the two IRAC bands: 

\begin{eqnarray}
M_{[3.6]}=-2.332 (\pm 0.106) \log(P) - 1.054 (\pm 0.020) \\
M_{[4.5]}=-2.336 (\pm 0.105) \log(P) - 1.091 (\pm 0.020) 
\end{eqnarray}

An in-depth study of the MIR \rrl{} PL relations based on theoretical
models calculated for the IRAC bands will be presented in a
forthcoming paper.

\section{Conclusions}
\label{sec:concl}

We have presented new \rrl{} MIR PL relations for the nearby globular cluster M4. 
Accurate IRAC photometry allows us to reduce the error in the PL slope by 
a factor of at least two from previous works. We have also demonstrated that 
PL relations in the IRAC bands are consistent with previous calibrations using WISE
photometry. 

We also presented optical and infrared PC relations. The steeper slope and smaller
dispersion of the $V - [3.6]$ relation suggests that it could be preferred over $V -K$ 
for use in the Baade-Wesselink method. Unlike for Cepheid variables, the $[3.6] - [4.5]$ 
relation is very shallow, indicating that there is no CO absorption, and therefore little
metallicity effect, in the 4.5~\micron{} band. As a result, the 4.5~\micron{} band can be
used as a distance indicator. 

We calibrated the zero point of the MIR PL relations with five Galactic \rrl{} stars with known
distances. The uncertainty in their distances is the largest source of error, and we will
be able to provide a much more precise calibration using upcoming results from the 
Gaia mission. Due to the uncertainty of the available distances to Galactic \rrl{} stars and 
the influence of metallicity in the PL relation, we provided a separate PL relation, 
where the zero point is calibrated using a distance modulus to M4 based on 
theoretical NIR PLZ relations. Future work will include analysis the remaining 
GCCs in the CRRP program in order to provide a comprehensive calibration of the 
\rrl{} PL relation. This calibration can then be compared to other distance indicators, such as Type II
Cepheids and Delta Scuti variables. 

\acknowledgments

We would like to thank the anonymous referee for providing constructive comments to improve the content of this paper. This work is based [in part] on observations made with the Spitzer Space Telescope, which is operated by the Jet Propulsion Laboratory, California Institute of Technology under a contract with NASA. This work was partially supported by PRIN-INAF 2011 ``Tracing the formation
and evolution of the Galactic halo with VST" (P.I.: M. Marconi) and by
PRIN-MIUR (2010LY5N2T) ``Chemical and dynamical evolution of the Milky Way
and Local Group galaxies" (P.I.: F. Matteucci).



{\it Facilities:} \facility{Spitzer (IRAC)}.

\clearpage



\begin{figure*}
\figurenum{4 cont}
\includegraphics[angle=0,scale=.9]{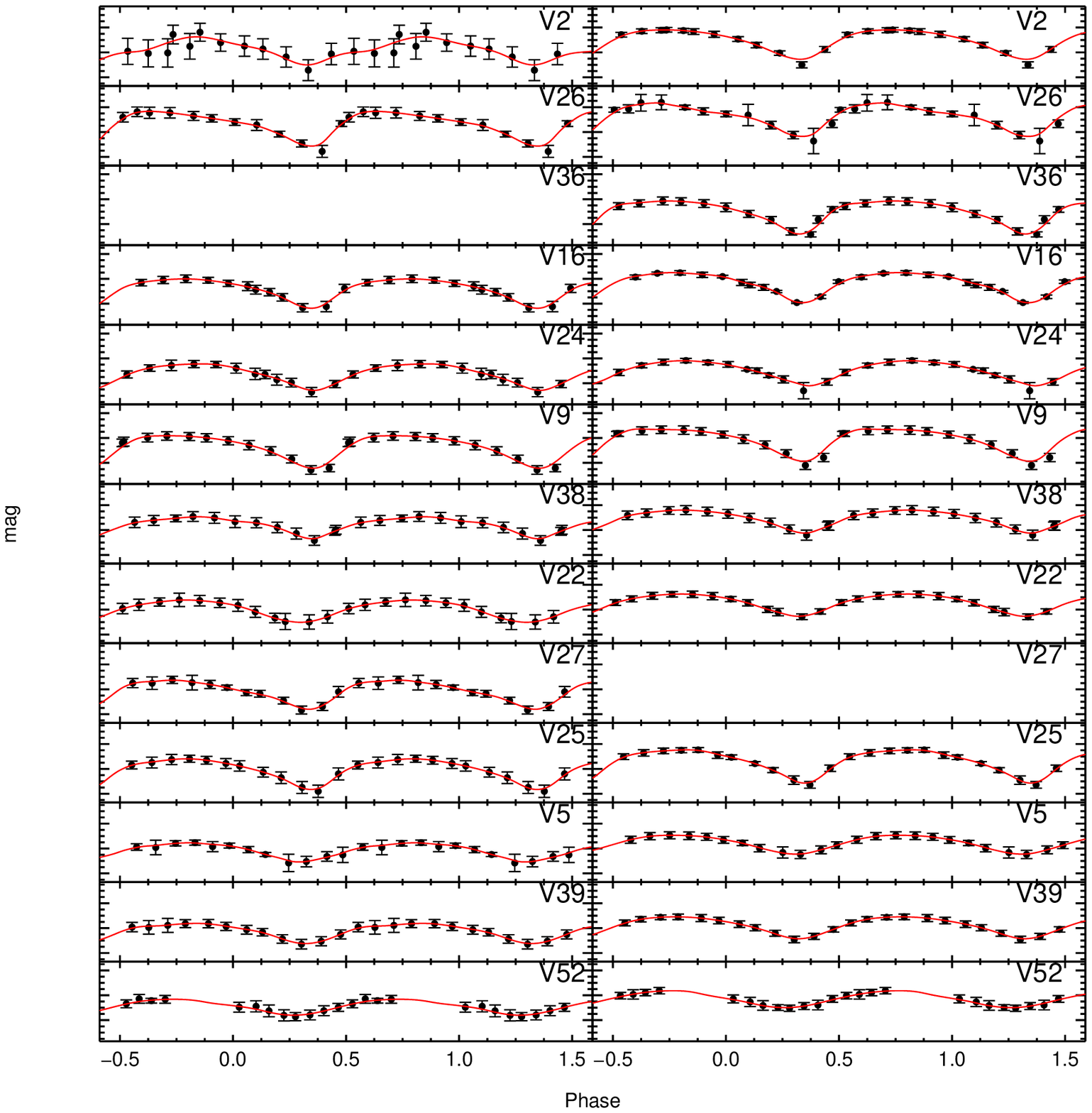}
\caption{Same as Figure~\ref{fig:FO-lightcurves} but for all FU
  pulsators observed by IRAC. }\label{fig:FU-lightcurves2}
\end{figure*}

\begin{figure*}
\includegraphics[angle=0,scale=0.9]{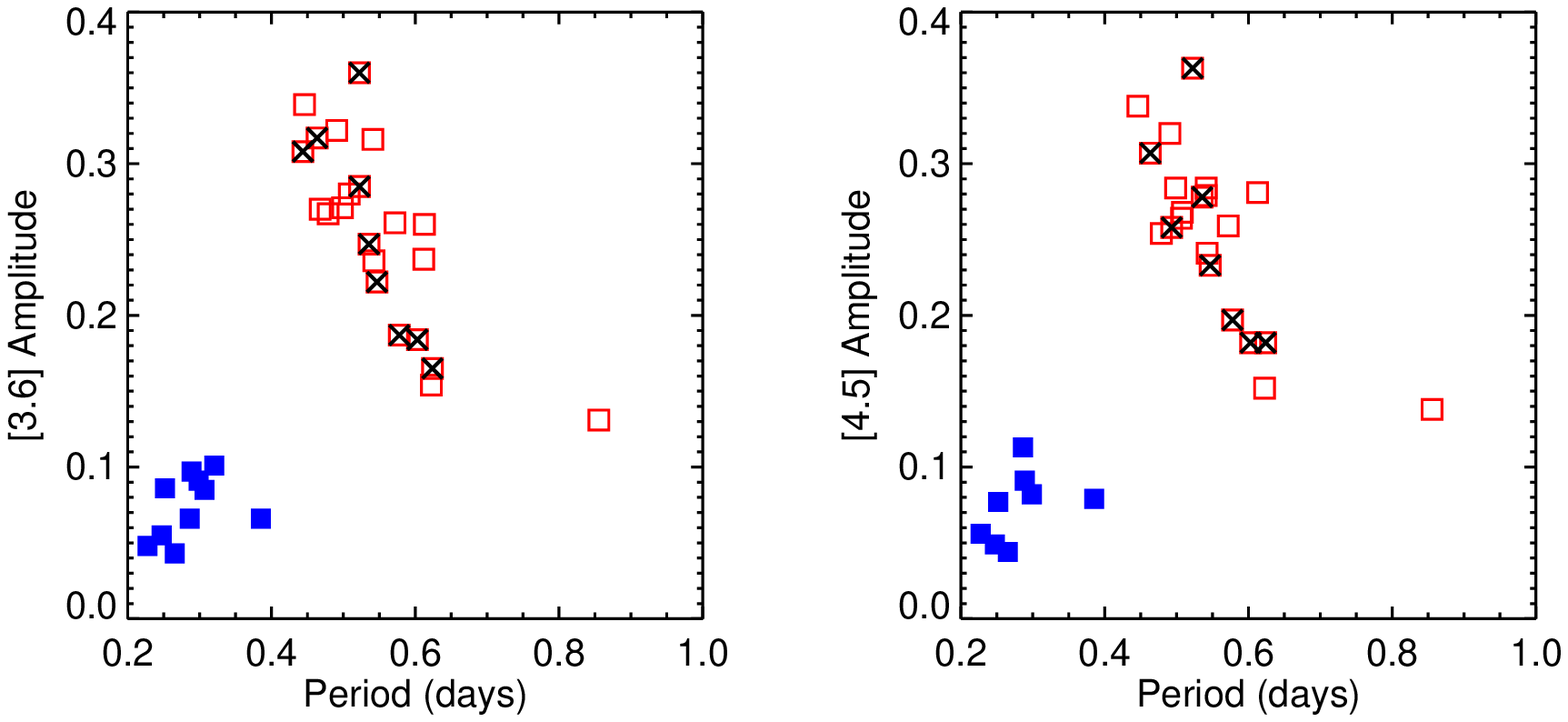}
\caption{Plot of the IRAC band amplitudes versus period for \rrl{} stars in
  M4 (Bailey diagram). First overtone and fundamental variables are
  represented by filled blue and open red squares. Candidate Blazhko stars are
  marked by black crosses. }\label{fig:bailey} 
\end{figure*}

\begin{figure*}
\includegraphics[angle=0,scale=0.9]{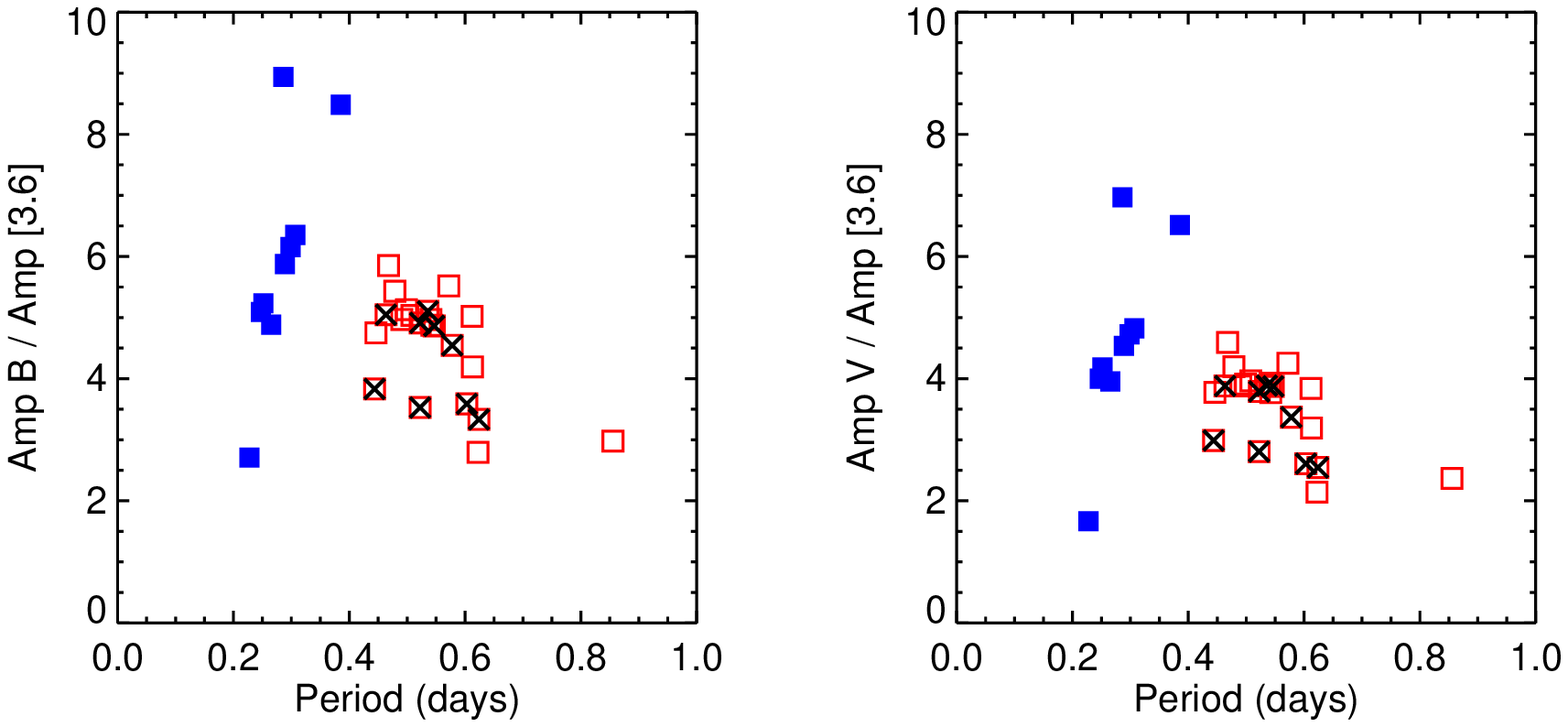}
\caption{Ratio between the optical and MIR band amplitudes for the
  first overtone (blue filled squares) and fundamental (red open squares)
  variables as a function of period. The left plot shows the ratio of
  the B to [3.6] amplitude, while the right plot shows the ratio of V
  band to [3.6] amplitude. Black crosses identify candidate Blazhko
  variables. The two FO \rrl{} stars (V40 and C1) that show an unusually high 
  optical to MIR ratio, could possibly indicate of the presence of a faint 
  optically-bright (hot) companion or blend. The FO \rrl{} (V49) with an 
  unusually low amplitude ratio, could indicate an infrared-bright (cool) 
  companion (or blend)}\label{fig:AP}  
\end{figure*}

\begin{figure*}
\includegraphics[angle=0,scale=0.8]{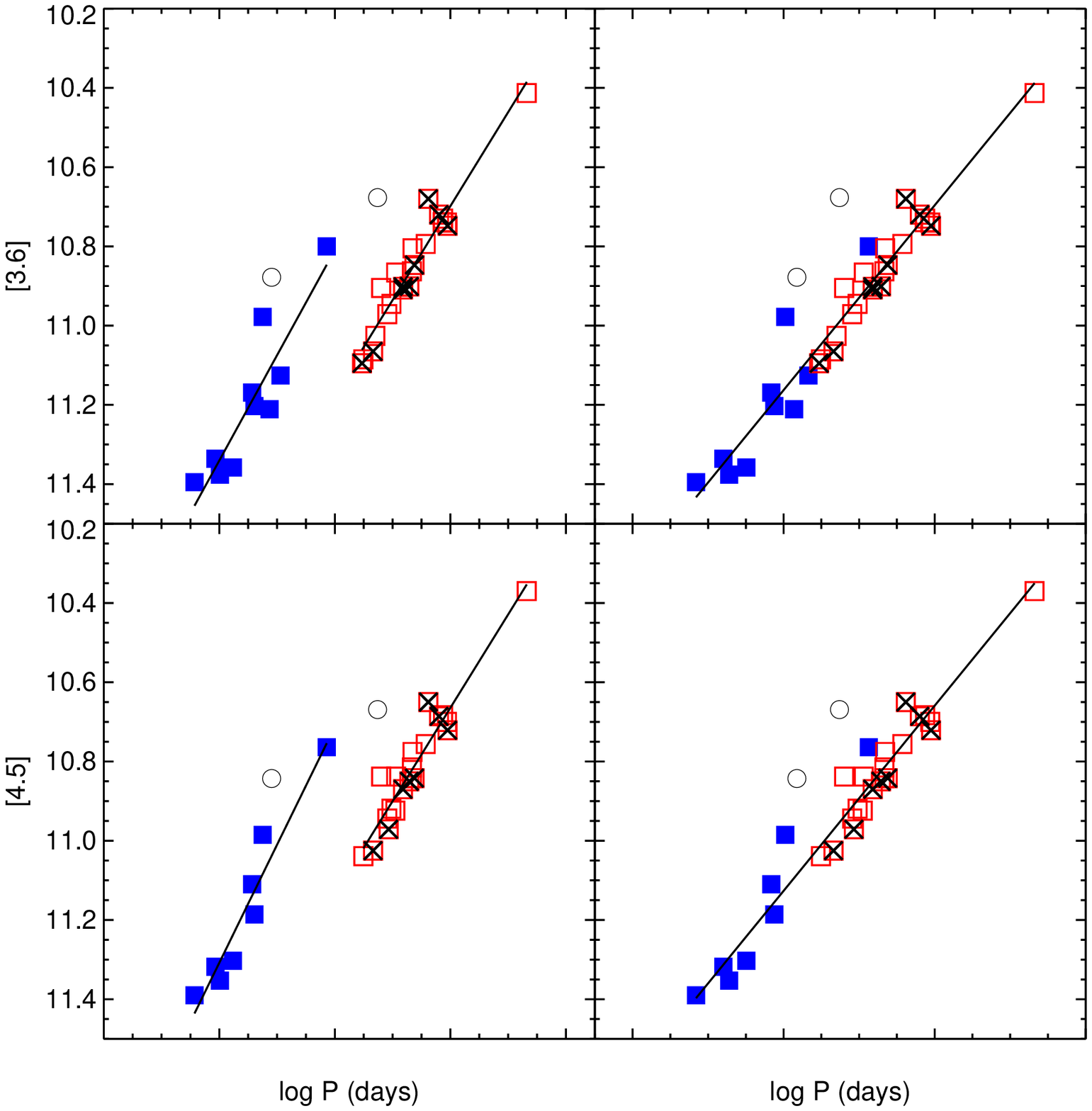}
\caption{IRAC period-luminosity relations for M4 \rrl{} stars. \emph{Left}
  --- FO (blue filled squares) and FU (red open squares) variables are fit
  independently (black solid lines). \emph{Right} --- single fit with
  fundamentalized FO \rrl{} stars. All magnitudes are corrected for extinction. 
  Random uncertainties in the average magnitudes
  of all variables are smaller than the symbols used, and are not
  shown. Candidate Blazhko stars are marked with black crosses. The open 
  circles mark the positions of the two stars (V20 and V21) that were not included in
  the fit of the PL relation due to blends with nearby sources. 
  }\label{fig:PL-warm} 
\end{figure*}

\begin{figure*}
\includegraphics[angle=0,scale=.9]{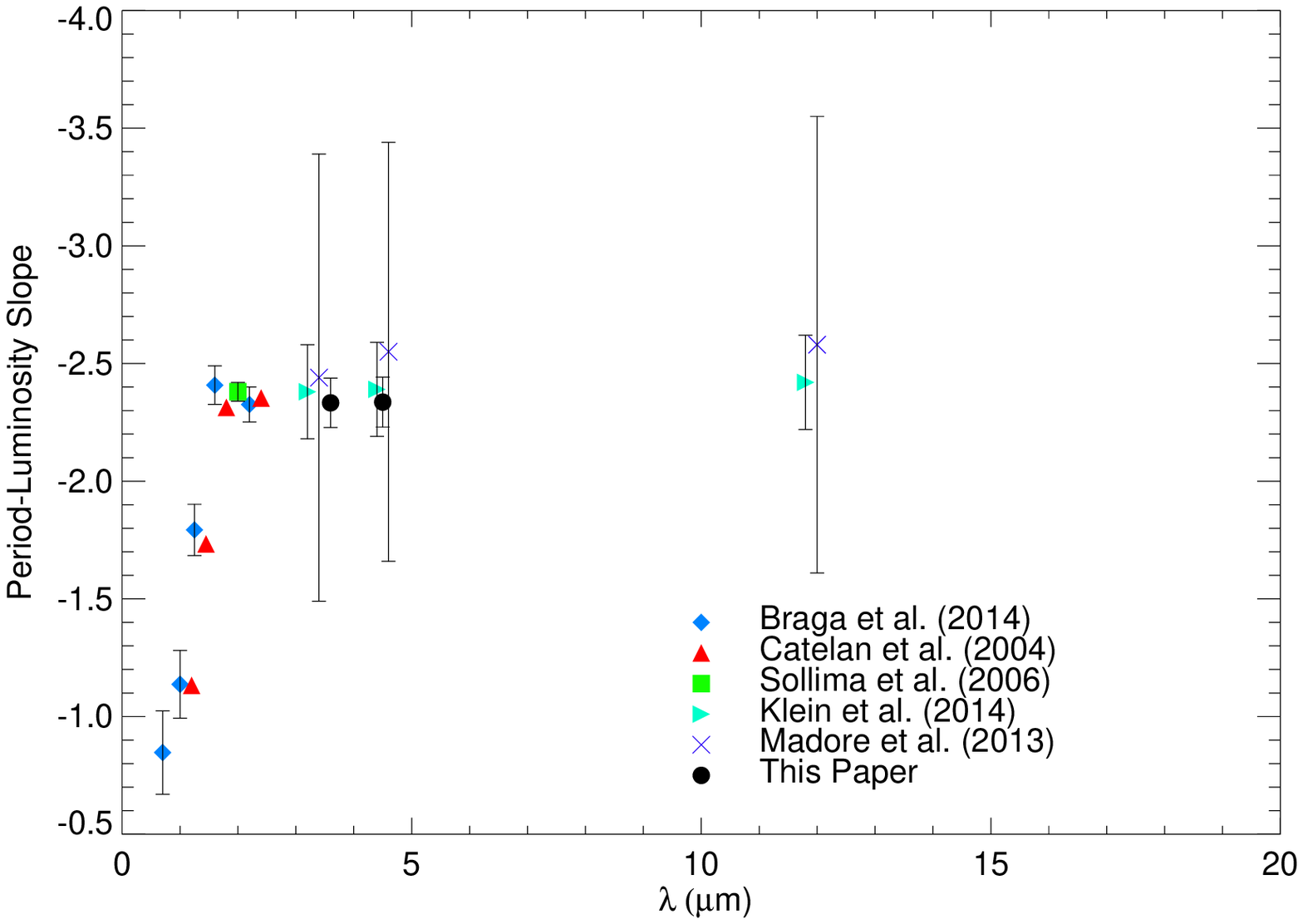}
\caption{An updated version of Figure 4 from \citet{mad13} showing the
  increase of the slope of the \rrl{} period-luminosity relation with
  wavelength. The optical and near-IR slopes are taken from
  \protect\citet{cat04}, \protect\citet{sol06} and \protect\citet{bra15}. The MIR slopes from \protect\citet{kle14}, \citet{mad13} 
  are marginally steeper than the values derived in this paper (using IRAC data only), but consistent within their uncertainties . For clarity, results from the same band are
  offset by 0.2~\micron{} (Catelan $+ 0.2$, Sollima $- 0.2$). Error
  bars are shown for all points, with the exception the theoretical
  prediction from \protect\citet{cat04}.}\label{fig:PL-parameters} 
\end{figure*}

\begin{figure*}
\includegraphics[angle=0,scale=.9]{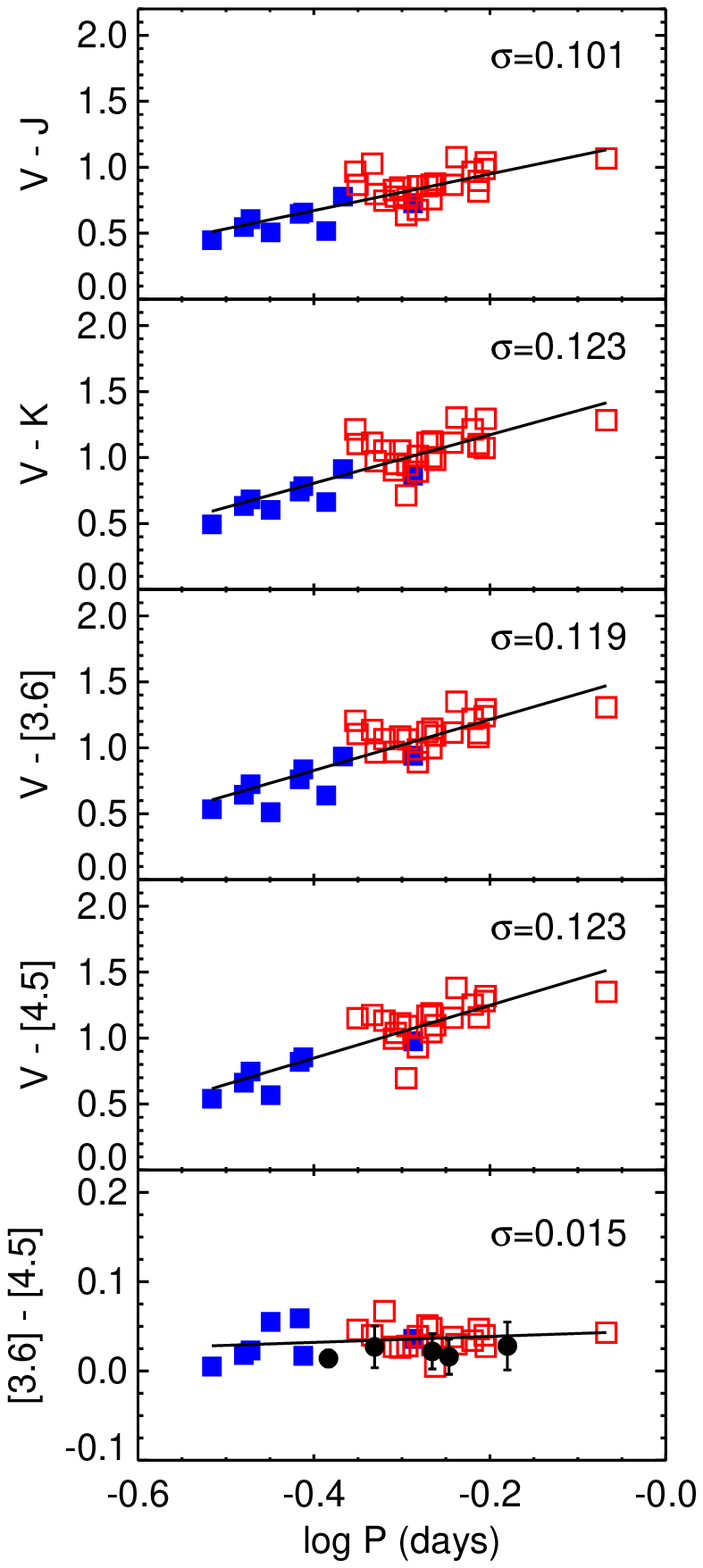}
\caption{The period-color relationship for the \rrl{} stars in M4. FO and FU
  variables observed with IRAC are shown in filled blue and open red squares
  respectively, with fundamentalized periods for the FO \rrl{}. Optical
  and NIR photometry is derived from \protect\citet{ste14}. The solid black line 
  represents the unweighted least squares fit. Five Galactic
  \rrl{} stars used for calibration are shown in the bottom panel as black
  circles (not included in the fit). All magnitudes are corrected for extinction. 
  Note all panels except the bottom span the same range in magnitude. }\label{fig:PC} 
\end{figure*}

\begin{figure*}
\includegraphics[angle=0,scale=.9]{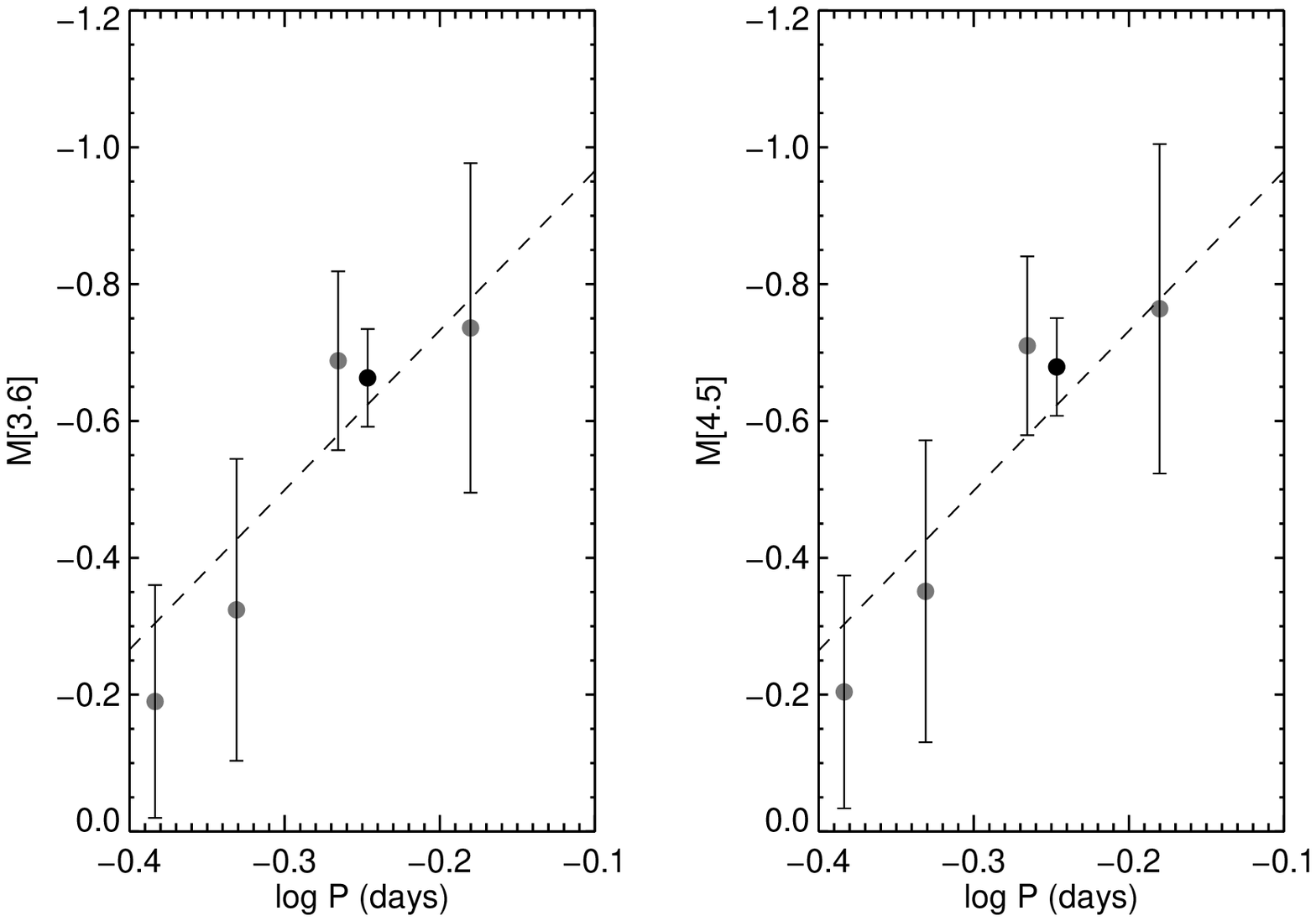}
\caption{The calibrated IRAC \rrl{} period-luminosity relation. The
  filled circles represent the absolute magnitudes of five Galactic
  \rrl{} stars. In order of increasing period: RZ Cep, XZ Cyg, UV Oct, RR
  Lyr (black point), and SU Dra. The period of RZ Cep has been
  fundamentalized with the same relationship adopted for the M4 FO
  \rrl{} stars: $\log P_{FU} = \log P_{FO} + 0.127$. The PL relations are
  shown by the dashed lines, with a slope determined by the M4 cluster
  data. }\label{fig:HST-fit}
\end{figure*}

\begin{figure*}
\includegraphics[angle=0,scale=0.9]{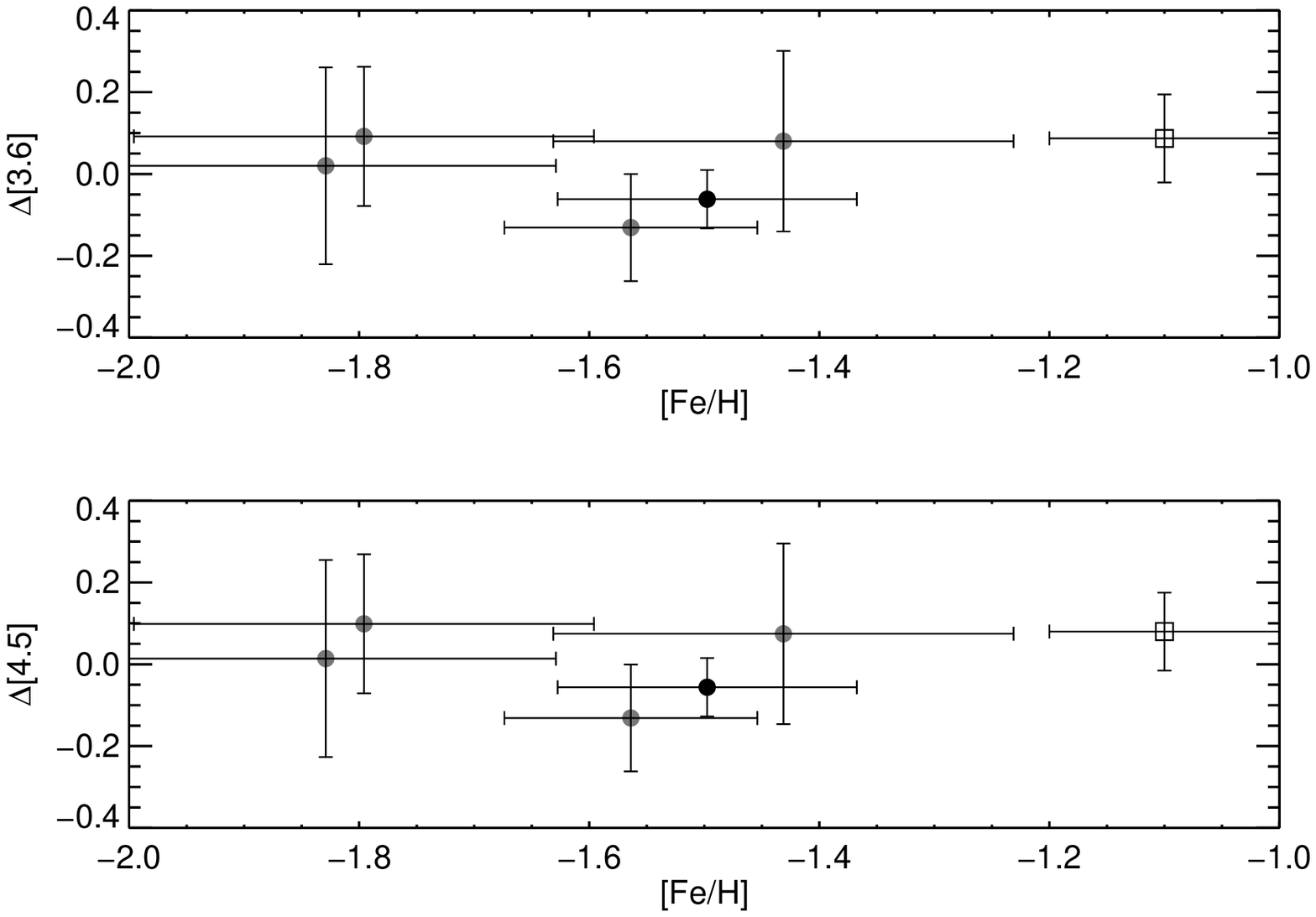}
\caption{Deviations from the period-luminosity relation as a function
  of metallicity. Galactic \rrl{} stars are shown with solid circles (the
  darker point, with the smallest error bars is RR~Lyr) and the average
  of all stars in M4 is shown as a open square. }\label{fig:met}
\end{figure*}

\clearpage 
\capstartfalse

\begin{deluxetable*}{cccccl}
\tabletypesize{\scriptsize}
\tablecaption{Photometry for V1. \label{tab:phot}}
\tablewidth{0pt}
\tablehead{
\colhead{MJD\tablenotemark{a}} & \colhead{[3.6] mag} & \colhead{$\sigma_{[3.6]}$} & \colhead{MJD\tablenotemark{a}} & \colhead{[4.5] mag} & \colhead{$\sigma_{[4.5]}$}  
} 
\startdata
  56446.223  & 11.354  & 0.030 & 56446.227 & 11.267 &  0.079 \\
  56446.285  & 11.376  & 0.010 & 56446.293 & 11.317  & 0.050 \\
  56446.336  & 11.394  & 0.003 & 56446.340 & 11.318  & 0.051 \\ 
  56446.391  & 11.329  & 0.011 & 56446.398 & 11.261  & 0.074 \\
  56446.438  & 11.302  & 0.012 & 56446.441 & 11.224  & 0.089 \\
  56446.492  & 11.312  & 0.021 & 56446.496 & 11.248  & 0.076 \\
  56446.543  & 11.358  & 0.016 & 56446.547 & 11.285  & 0.069 \\
  56446.598  & 11.385  & 0.035 & 56446.602 & 11.312  & 0.075 \\
  56446.652  & 11.344  & 0.021 & 56446.656 & 11.268 &  0.078 \\
  56446.699  & 11.309  & 0.015 & 56446.703 & 11.240  & 0.063 \\
  56446.746  & 11.295  & 0.029 & 56446.750 & 11.233  & 0.077 \\
  56446.797  & 11.340  & 0.023 & 56446.805 & 11.278  & 0.080 \\
\enddata
\tablenotetext{a}{MJD = JD - 2400000.5 days}
\tablecomments{Table 1 is published in its entirety in the electronic edition. A portion is shown here for guidance regarding its form and content.}
\end{deluxetable*}

\capstarttrue

\capstartfalse
\begin{deluxetable*}{lccccccccl}
\tabletypesize{\scriptsize}
\tablecaption{Mean IRAC magnitudes and amplitudes for the \rrl{} stars in M4. \label{tab:Smean}}
\tablewidth{0pt}
\tablehead{
\colhead{ID\tablenotemark{a}} & \colhead{$\alpha$ (J2000.0)} &  \colhead{Period (days)} & \colhead{$\log(P)$}& \colhead{[3.6] mag} & \colhead{[4.5] mag}  & \colhead{Amp$_{[3.6]}$} & \colhead{Amp$_{[4.5]}$} & \colhead{Mode\tablenotemark{b}} \\
 & \colhead{$\delta$ (J2000.0)} & & & & & & & 
}
\startdata
V1	&	16	23	13.67	&	0.28888261	&	-0.5392786	&	$11.278\pm0.007$	&	$11.244\pm0.021$\tablenotemark{c}	&	$0.10\pm0.01$	&	$0.09\pm0.04$	&	RRc	\\
        & 	-26	30	53.6 && & & & & & \\
V2	&	16	23	16.30	&	0.5356819	&	-0.271093	&	$10.976\pm0.027$\tablenotemark{c}	&	$10.908\pm0.010$	&	$0.25\pm0.06$	&	$0.28\pm0.02$	&	RRab*	\\
	&	-26	34	46.6	&& & & & & & \\
V3	&	16	23	19.47	&	0.50667787	&	-0.2952681	&	\nodata	&	$10.982\pm0.008$	&	\nodata	&	$0.26\pm0.01$	&	RRab	\\
	&	-26	39	57.0	&& & & & & & \\
V5	&	16	23	21.03	&	0.62240112	&	-0.2059296	&	$10.815\pm0.012$	&	$10.758\pm0.009$	&	$0.15\pm0.04$	&	$0.15\pm0.02$	&	RRab	\\
	&	-26	33	05.8	&& & & & & & \\
V6	&	16	23	25.79	&	0.3205151	&	-0.4941515	&	$11.201\pm0.013$	&	\nodata	&	$0.10\pm0.02$	&	\nodata	&	RRc	\\
	&	-26	26	16.3	&& & & & & & \\
V7	&	16	23	25.95	&	0.49878722	&	-0.3020847	&	$11.020\pm0.013$	&	$10.977\pm0.010$	&	$0.27\pm0.02$	&	$0.28\pm0.01$	&	RRab	\\
	&	-26	27	41.9	&& & & & & & \\
V8	&	16	23	26.16	&	0.50822359	&	-0.2939452	&	$10.941\pm0.013$	&	$10.896\pm0.009$	&	$0.28\pm0.02$	&	$0.27\pm0.01$	&	RRab	\\
	&	-26	29	41.6	&& & & & & & \\
V9	&	16	23	26.80	&	0.57189447	&	-0.2426841	&	$10.869\pm0.012$	&	$10.814\pm0.011$	&	$0.26\pm0.02$	&	$0.26\pm0.02$	&	RRab	\\
	&	-26	29	48.0	&& & & & & & \\
V10	&	16	23	29.21	&	0.49071753	&	-0.3091684	&	$11.046\pm0.015$	&	$11.002\pm0.011$	&	$0.32\pm0.03$	&	$0.32\pm0.01$	&	RRab	\\
	& 	-26	28	54.3	&& & & & & & \\
V11	&	16	23	29.98	&	0.49320868	&	-0.3069693	&	\nodata	&	$11.029\pm0.010$	&	\nodata	&	$0.26\pm0.02$	&	RRab*	\\
	&	-26	36	24.5	& & & & & & & \\
V12	&	16	23	30.82	&	0.4461098	&	-0.3505582	&	$11.160\pm0.023$	&	$11.097\pm0.013$	&	$0.34\pm0.04$	&	$0.34\pm0.01$	&	RRab	\\
	&	-26	34	58.9	& & & & & & & \\
V14	&	16	23	31.29	&	0.46353111	&	-0.3339211	&	$11.140\pm0.017$	&	$11.083\pm0.014$	&	$0.32\pm0.05$	&	$0.31\pm0.04$	&	RRab*	\\
	& 	-26	35	34.5	&& & & & & & \\
V15	&	16	23	31.96	&	0.44366077	&	-0.352949	&	$11.170\pm0.014$	&	\nodata	&	$0.31\pm0.02$	&	\nodata	&	RRab*	\\
	&	-26	24	18.1	& & & & & & & \\
V16	&	16	23	32.50	&	0.54254824	&	-0.2655616	&	$10.880\pm0.011$	&	$10.834\pm0.008$	&	$0.24\pm0.03$	&	$0.24\pm0.01$	&	RRab	\\
	&	-26	30	23.0	& & & & & & & \\
V18	&	16	23	34.70	&	0.47879201	&	-0.3198531	&	$10.980\pm0.016$	&	$10.896\pm0.014$	&	$0.27\pm0.04$	&	$0.25\pm0.03$	&	RRab	\\
	& 	-26	31	04.6	&& & & & & & \\
V19	&	16	23	35.05	&	0.46781108	&	-0.3299295	&	$11.101\pm0.014$	&	\nodata	&	$0.27\pm0.03$	&	\nodata	&	RRab	\\
	& 	-26	25	36.3	&& & & & & & \\
V20	&	16	23	35.39	&	0.30941948	&	-0.5094523	&	$10.953\pm0.031$	&	$10.901\pm0.031$	&	$0.09\pm0.06$	&	$0.06\pm0.06$	&	RRc	\\
	&	-26	32	35.9	& & & & & & & \\
V21	&	16	23	35.93	&	0.47200742	&	-0.3260512	&	$10.752\pm0.016$	&	$10.727\pm0.016$	&	$0.22\pm0.04$	&	$0.24\pm0.03$	&	RRab	\\
	& 	-26	31	33.6	&& & & & & & \\
V22	&	16	23	36.95	&	0.60306358	&	-0.2196369	&	$10.795\pm0.014$	&	$10.744\pm0.009$	&	$0.18\pm0.04$	&	$0.18\pm0.02$	&	RRab*	\\
	& 	-26	30	13.0	&& & & & & & \\
V23	&	16	23	37.33	&	0.29861557	&	-0.5248876	&	$11.053\pm0.011$	&	$11.043\pm0.013$	&	$0.09\pm0.02$	&	$0.08\pm0.04$	&	RRc	\\
	& 	-26	31	56.1	&& & & & & & \\
V24	&	16	23	38.04	&	0.54678333	&	-0.2621847	&	$10.922\pm0.012$	&	$10.900\pm0.010$	&	$0.22\pm0.02$	&	$0.23\pm0.03$	&	RRab*	\\
	&	-26	30	41.8	& & & & & & & \\
V25	&	16	23	39.42	&	0.61273479	&	-0.2127275	&	$10.805\pm0.013$	&	$10.741\pm0.010$	&	$0.26\pm0.04$	&	$0.28\pm0.02$	&	RRab	\\
	&	-26	30	21.3	& & & & & & & \\
V26	&	16	23	41.65	&	0.54121739	&	-0.2666283	&	$10.938\pm0.013$	&	$10.873\pm0.017$	&	$0.32\pm0.02$	&	$0.28\pm0.04$	&	RRab	\\
	&	-26	32	41.1	& & & & & & & \\
V27	&	16	23	43.17	&	0.61201829	&	-0.2132356	&	$10.814\pm0.012$	&	\nodata	&	$0.24\pm0.03$	&	\nodata	&	RRab	\\
	& 	-26	27	16.3	&& & & & & & \\
V28	&	16	23	53.60	&	0.52234107	&	-0.2820458	&	$10.984\pm0.013$	&	$10.928\pm0.014$	&	$0.36\pm0.02$	&	$0.36\pm0.02$	&	RRab*	\\
	& 	-26	30	5.30	&& & & & & & \\
V29	&	16	23	58.25	&	0.52248466	&	-0.2819265	&	$10.977\pm0.013$\tablenotemark{c}	&	\nodata	&	$0.29\pm0.03$	&	\nodata	&	RRab*	\\
	& 	-26	21	35.1	&& & & & & & \\
V36	&	16	23	19.45	&	0.54130918	&	-0.2665546	&	\nodata	&	$10.900\pm0.011$	&	\nodata	&	$0.28\pm0.03$	&	RRab	\\
	& 	-26	35	49.0	&& & & & & & \\
V37	&	16	23	31.60	&	0.24734353	&	-0.6066994	&	$11.411\pm0.013$	&	$11.376\pm0.007$	&	$0.06\pm0.03$	&	$0.05\pm0.01$	&	RRc	\\
	& 	-26	31	30.6	&& & & & & & \\
V38	&	16	23	32.87	&	0.57784635	&	-0.2381876	&	$10.755\pm0.012$	&	$10.708\pm0.011$	&	$0.19\pm0.03$	&	$0.20\pm0.03$	&	RRab*	\\
	& 	-26	33	03.5	&& & & & & & \\
V39	&	16	23	34.67	&	0.623954	&	-0.2048474	&	$10.823\pm0.012$	&	$10.779\pm0.008$	&	$0.17\pm0.03$	&	$0.18\pm0.02$	&	RRab*	\\
	& 	-26	32	52.1	&& & & & & & \\
V40	&	16	23	34.59	&	0.38533005	&	-0.4141671	&	$10.875\pm0.020$	&	$10.822\pm0.011$	&	$0.07\pm0.04$	&	$0.08\pm0.02$	&	RRc	\\
	& 	-26	30	44.2	&& & & & & & \\
V41	&	16	23	39.50	&	0.2517418	&	-0.5990447	&	$11.451\pm0.011$	&	$11.411\pm0.007$	&	$0.09\pm0.03$	&	$0.08\pm0.01$	&	RRc	\\
	& 	-26	33	59.8	&& & & & & & \\
V42	&	16	24	02.00	&	0.3068549	&	-0.5130669	&	$11.286\pm0.032$\tablenotemark{c}	&	\nodata	&	$0.09\pm0.06$	&	\nodata	&	RRc	\\
	& 	-26	22	13.0	&& & & & & & \\
V49	&	16	23	45.25	&	0.22754331	&	-0.6429359	&	$11.470\pm0.011$	&	$11.448\pm0.010$	&	$0.05\pm0.03$	&	$0.06\pm0.02$	&	RRc	\\
	& 	-26	31	28.4	&& & & & & & \\
V52	&	16	23	24.06	&	0.85549784	&	-0.0677811	&	$10.488\pm0.015$	&	$10.428\pm0.015$	&	$0.13\pm0.02$	&	$0.14\pm0.02$	&	RRab	\\
	&	-26	30	27.8	& & & & & & & \\
V61	&	16	23	29.76	&	0.26528645	&	-0.5762849	&	$11.433\pm0.015$	&	$11.361\pm0.006$	&	$0.04\pm0.03$	&	$0.04\pm0.01$	&	RRc	\\
	& 	-26	29	50.3	&& & & & & & \\
C1	&	16	23	36.43	&	0.2862573	&	-0.5432434	&	$11.244\pm0.029$	&	$11.168\pm0.032$	&	$0.07\pm0.06$	&	$0.11\pm0.06$	&	RRc	\\
	&	-26	30	42.8	& & & & & & & \\
\enddata
\tablenotetext{a}{The ID given in Clement's catalog, with the exception of C1, newly named in \citet{ste14}}
\tablenotetext{b}{The pulsation mode (RRab=FU, RRc=FO). Candidate Blazhko stars are indicated by an asterisk.}
\tablenotetext{c}{The uncertainty is derived using the repeatability parameter given in DAOPHOT.}
\end{deluxetable*}
\capstarttrue

\capstartfalse
\begin{deluxetable*}{lccccccccc}
\tabletypesize{\scriptsize}
\tablecaption{Observed MIR period-luminosity relations for \rrl{} stars in M4. \label{tab:PL}}
\tablewidth{0pt}
\tablehead{
\colhead{Band} & \colhead{a\tablenotemark{a}} & \colhead{b\tablenotemark{a}}  & \colhead{$\sigma$} & \colhead{a\tablenotemark{b}} & \colhead{b\tablenotemark{b}} & \colhead{$\sigma$} 
& \colhead{a\tablenotemark{c}} & \colhead{b\tablenotemark{c}} & \colhead{$\sigma$} 
}
\startdata
            &                     &   FO             &           &                     &       FU          &            &                     &  FU+FO          &          \\      
$[3.6]$ & 11.207         & -2.658         &  0.079 & 10.841         & -2.370          & 0.040   & 10.929        & -2.332            & 0.056 \\  
            & $\pm0.027$ & $\pm0.428$ &           & $\pm0.009$ & $\pm0.139$  &            & $\pm0.010$ & $ \pm 0.106$ &         \\
            &                     &                     &           &                     &                      &             &                    &                       &          \\
$[4.5]$ & 11.159           & -2.979          & 0.057 & 10.806         & -2.355           & 0.045  & 10.893         & -2.336           & 0.054 \\
            & $\pm0.022$ & $\pm0.337$   &           & $\pm0.010$ & $\pm0.168$   &           & $\pm0.010$ & $\pm 0.105$  &    \\
            &                      &                    &            &                     &                      &           &                      &                        &     \\
\enddata
\tablenotetext{a}{PL parameters of the form $m = a + b(\log(P)+0.55)$}
\tablenotetext{b}{PL parameters of the form $m = a + b(\log(P)+0.26)$}
\tablenotetext{c}{PL parameters of the form $m = a + b(\log(P)+0.30)$}
\end{deluxetable*}
\capstarttrue

\capstartfalse
\begin{deluxetable*}{lccccc}
\tabletypesize{\scriptsize}
\tablecaption{Fundamentalized period-color relations. \label{tab:PC}}
\tablewidth{0pt}
\tablehead{
\colhead{Color} & \colhead{a\tablenotemark{a}} &  \colhead{b\tablenotemark{a}} & \colhead{$\sigma$} 
}
\startdata
V - J & $0.808\pm0.018$ & $1.421\pm0.192$ & 0.102 \\
V - K & $0.985\pm0.021$ & $1.874\pm0.232$ & 0.123 \\
V - [3.6] & $1.021\pm0.021$ & $1.934\pm0.221$ & 0.117 \\
V - [4.5] & $1.047\pm0.023$ & $2.012\pm0.238$ & 0.121 \\
$[3.6] - [4.5]$ & $0.034\pm0.003$ & $0.046\pm0.033$ & 0.017 \\
\enddata
\tablenotetext{a}{PC parameters of the form $color= a + b(\log(P)+0.3)$}
\end{deluxetable*}
\capstarttrue

\capstartfalse
\begin{deluxetable*}{lccccc}
\tabletypesize{\scriptsize}
\tablecaption{IRAC magnitudes for Galactic \rrl{} stars. \label{tab:gal}}
\tablewidth{0pt}
\tablehead{
\colhead{} & \colhead{RZ Cep} & \colhead{XZ Cyg}  & \colhead{UV Oct} & \colhead{RR Lyr} & \colhead{SU Dra} 
}
\startdata
Period & 0.308645 & 0.466579 & 0.542600263 & 0.566805 & 0.660419 \\
$[Fe/H]$ & $-1.80\pm0.2$ & $-1.43\pm0.2$ & $-1.56\pm0.11$ & $-1.50\pm0.13$ & $-1.83\pm0.2$ \\
$E(B-V)$    & 0.252               &  0.100                & 0.090                &  0.042               &  0.010       \\
$A_{[3.6]}$ & 0.051               & 0.020                 &  0.018               & 0.009                & 0.002                 \\
$A_{[4.5]}$ & 0.039               & 0.016                &  0.014               & 0.007                & 0.002                 \\
                   &                          &                         &                           &                          &                    \\
Parallax (mas) & $2.54\pm0.19$\tablenotemark{a} & $1.67\pm0.17$ & $1.71\pm0.10$ & $3.77\pm0.13$ & $1.42\pm0.16$ \\
LKH correction & -0.05 & -0.09 & -0.03 & -0.02 & -0.11 \\
$(m-M)_0$\tablenotemark{b} & $8.03\pm0.16$ & $8.98\pm0.22$ & $8.87\pm0.13$ & $7.14\pm0.07$ & $9.35\pm0.24$ \\
                  &                          &                         &                           &                          &                    \\
$[3.6]$     & $7.891\pm0.007$ & $8.676\pm0.016$ & $8.200\pm0.014$ & $6.486\pm0.014$& $8.616\pm0.019$ \\
$[4.5]$     & $7.865\pm0.006$& $8.645\pm0.017$& $8.174\pm0.014$ & $6.468\pm0.014$  & $8.588\pm0.019$  \\
                   &                          &                           &                         &                              &                     \\
$M[3.6]$ & $-0.190\pm0.160$ & $-0.324\pm0.221$ & $-0.688\pm0.131$ & $-0.663\pm0.071$ & $-0.736\pm0.241$  \\
$M[4.5]$ & $-0.204\pm0.160$ & $-0.361\pm0.221$ & $-0.710\pm0.131$ & $-0.679\pm0.071$ & $-0.764\pm0.241$  \\                
\enddata
\tablenotetext{a}{\protect\citet{ben11} provides two different parallax values for this star. We adopted the value given in their Section 4.3.2, rather than 
the number listed in their Table 8, which appears to be inconsistent with the PL relation.}
\tablenotetext{b}{Our distance moduli are slightly different than the values in Table 8 of \protect\citet{ben11} due to typographical 
errors in that paper.}

\end{deluxetable*}
\capstarttrue

\end{document}